# Tackling the Crowdsourced Shared-Trip Delivery Problem at Scale with a Novel Decomposition Heuristic

**Dingtong Yang**[a, b]
E-mail: dingtony@uci.edu
ORCiD: 0000-0001-7377-4531

**Michael F. Hyland**[a, b]**, Corresponding Author**
Email: hylandm@uci.edu
Phone: (949) 824-5084
ORCiD: 0000-0001-8394-8064
Institute of Transportation Studies, 4000 Anteater Instruction and Research Bldg. (AIRB)
Irvine, CA 92697-3600

**R. Jayakrishnan**[a, b]
E-mail: rjayakri@uci.edu

[a] University of California-Irvine, Civil and Environmental Engineering
[b] University of California-Irvine, Institute of Transportation Studies

## Abstract

This paper presents a set-partitioning formulation and a novel decomposition heuristic (D-H) solution algorithm to solve large-scale instances of the urban crowdsourced shared-trip delivery (CSD) problem. The CSD problem involves dedicated vehicles (DVs) and shared personal vehicles (SPVs) fulfilling delivery orders, wherein the SPVs have their own trip origins and destinations. The D-H begins by assigning as many package delivery orders (PDOs) to SPVs as possible, where the D-H enumerates the set of routes each SPV can feasibly traverse and then solves a PDO-SPV-route assignment problem. For PDO-DV assignment and DV routing, the D-H solves a multi-vehicle routing problem with time-window, tour duration, and capacity constraints using an insertion heuristic. Finally, the D-H seeks potential solution improvements by switching PDOs between SPV and DV routes through a simulated annealing (SA)-inspired procedure. The D-H outperforms a commercial solver in terms of computational efficiency while obtaining near-optimal solutions for small problem instances. The SA-inspired switching procedure outperforms a large neighborhood search algorithm regarding run time, and the two are comparable regarding solution quality. Finally, the paper uses the D-H to analyze the impact of several relevant factors on city-scale CSD system performance, namely the number of participating SPVs and the maximum willingness to detour of SPVs. Consistent with the existing literature, we find that CSD can substantially reduce delivery costs. However, we find that CSD can increase vehicle miles traveled. Our findings provide meaningful insights for logistics practitioners, while the algorithms illustrate promise for large real-world systems.

# 1 Introduction

To better accommodate the increasing demand for urban parcel delivery, several logistics companies now offer crowdsourced delivery (*CD*) services. In CD services, the logistics provider pays non-employee (also known as "gig" or "CD") drivers—usually in addition to fully employed drivers—to deliver packages (Rai et al., 2017). Typically, CD drivers use a vehicle that they own or lease. In contrast, drivers employed by logistics providers typically use a vehicle owned by the logistics provider dedicated to goods delivery.

Private firms and academic researchers are testing and evaluating various CD methods, respectively. Retailing, transportation, and logistics companies like Walmart Spark Driver, Amazon Flex, and Uber Eats have launched such programs. Alnaggar et al. (2021) provide a detailed summary of current CD programs. Researchers have also proposed various ways to formulate and solve logistics problems related to CD. Section 2 provides a detailed review of the current literature related to CD, also known as crowd-shipping.

In this paper, we focus on a specific type of CD service where package delivery orders (PDOs) "share rides" with CD drivers who have their own trips to complete. This problem is similar to several other CD logistical problems in the literature, including the "vehicle routing problem with occasional drivers (VRPOD)" in Archetti et al. (2016), the "crowdsourced delivery with ad hoc drivers" in Arslan et al. (2019), and the "crowdsourced delivery with in-store customers" in Dayarian and Savelsbergh (2020). The CD service we define is an extension of the service type in Dayarian and Savelsbergh (2020)—in both studies, potential CD drivers have their own trip origin and destination and can serve PDOs en-route between these two locations. However, unlike Dayarian and Savelsbergh (2020), we permit CD driver origins to include locations other than retail stores. Several other studies also consider CD drivers with flexible trip origins (Le et al., 2021; Mousavi et al., 2022; Nieto-Isaza et al., 2022). Not limiting the CD driver pool to just in-store customers should increase the market for CD drivers while keeping the core idea behind CD: utilizing space inside personal vehicles and taking advantage of the enormous number of daily personal vehicle trips. We name the specific problem we address, the crowdsourced shared-trip delivery (*CSD*) problem.

There are several names for CD drivers in the literature, including "occasional drivers" in Archetti et al. (2016, 2021) and Mohri et al. (2023) and "committed drivers" in Savelsbergh and Ulmer (2022) and Ulmer and Savelsbergh (2020). To further differentiate our CD drivers, whose primary purpose is not serving PDOs, from those CD or gig drivers who provide delivery services for a fixed period, we name our drivers crowdsourced shared-trip delivery (CSD) drivers. Moreover, we refer to their vehicles as shared personal vehicles (SPVs).

Figure 1 displays a diagram of the CSD system. A central manager is responsible for delivering PDOs from the distribution center to locations throughout the service area. The packages in PDOs are small- to medium-sized parcels (without hazardous material) that can easily fit inside the trunk or the seats of ordinary sedans. The PDOs each have a latest delivery time, distributed across the day, representing commitments made by the logistic provider or retail company to consumers. In the proposed system, SPVs travel to the distribution center, pick up PDOs, drop off the PDOs, and travel to their own destinations.

The SPV drivers are compensated based on the number of PDOs they deliver and their total detour distance. The central manager assigns PDOs to drivers. To make assignment decisions and compensate drivers for their detour distance, drivers must share their personal trip origin and destination with the platform. The central manager also has a fleet of trucks/vans (named "dedicated vehicles" and abbreviated as DVs) available to deliver parcels. The objective of the central manager is to minimize total delivery costs. The decision levers available to the central manager include the partitioning of PDOs between the SPVs and the DV fleet, the assignment of PDOs to specific SPVs and DVs, and the routing of each DV and SPV.

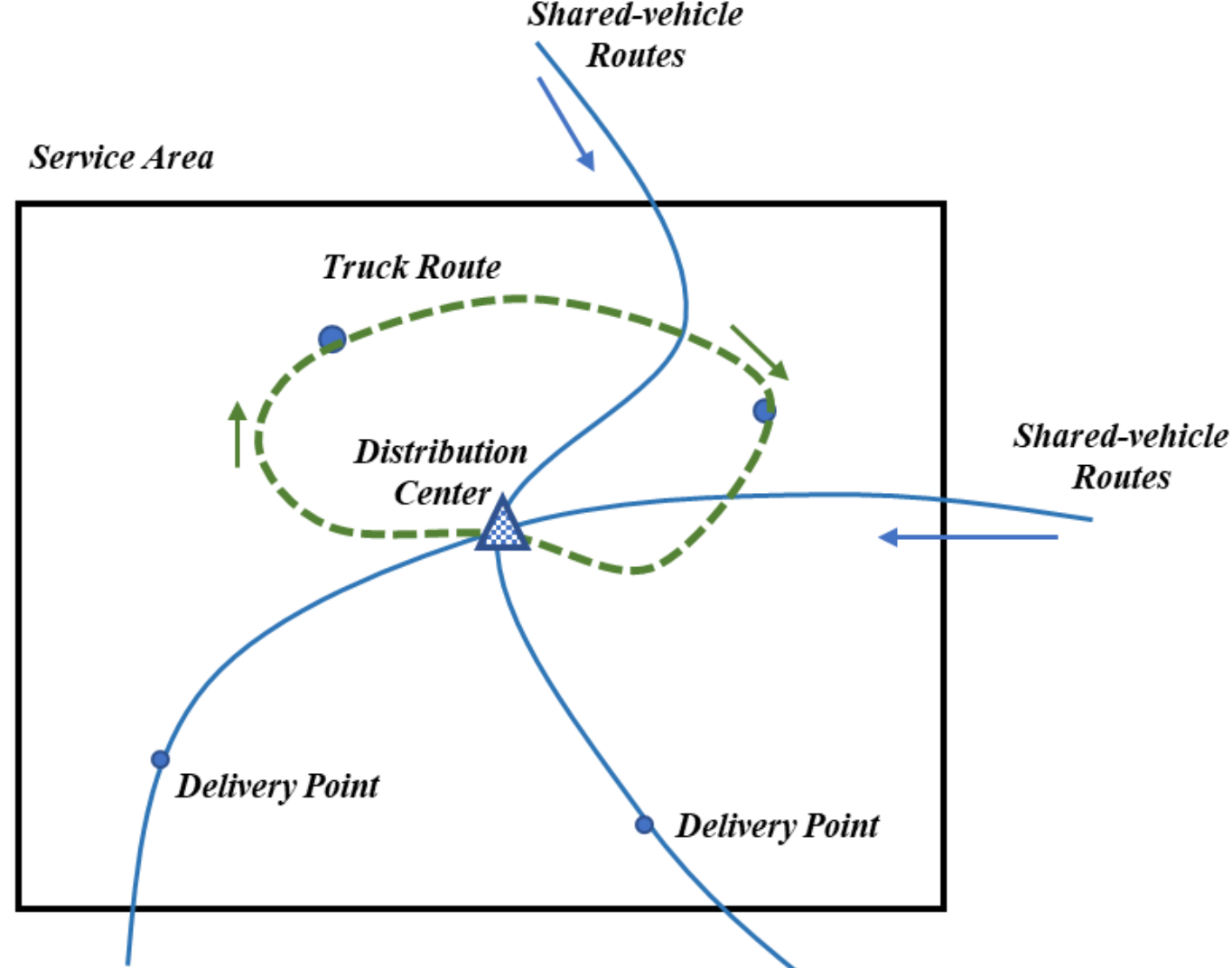

**Figure 1: Crowdsourced shared-trip delivery (CSD) system**

Several research groups also address the CSD problem (Archetti et al., 2016; Dahle et al., 2019; Kafle et al., 2017; Macrina et al., 2017, 2020). These studies find that sufficient driver participation is critical to the success of CD services. However, the problem instances in these prior studies only include 60-100 potential CSD drivers.

The overarching goal of this study is to develop a modeling approach and solution method to solve the CSD problem at scale. Additionally, we aim to provide insights into the potential benefits of CSD at scale. Specifically, we aim to address the CSD problem for cases with over 1000 CSD drivers. We also aim to generalize the existing CSD problem wherein only in-store customers can serve as CSD drivers, by allowing other drivers with their own trip origin and destination to serve as potential CSD drivers.

We propose a novel decomposition heuristic (D-H) to solve large-scale instances. The D-H handles the assignment of PDOs to SPVs and DVs separately but attempts to improve the solution by switching PDOs between SPVs and DVs. The switching procedure improves the solution quality iteratively through a simulated annealing (SA)-inspired procedure.

This paper makes several contributions to the academic literature and the practice of CD logistics.

1. (As stated before, handling a large number of SPVs (in the range of 1000) and efficiently tackling the CSD problem at scale are the major gaps in the literature. This study bridges the gap by proposing a D-H algorithm to solve large instances of the vehicle-PDO matching and routing problem for a mixed fleet of DVs and SPVs.
2. The second contribution lies in the design of the D-H algorithm. The algorithm provides a novel approach to handle the CSD problem. We first decompose the problem by vehicle type, where the two vehicle types are SPV and DV. We optimally assign PDOs to SPVs first; then, we assign the remaining PDOs to DVs using a standard insertion-based heuristic. We then iteratively improve the solution by swapping PDOs between SPVs and DVs—the next contribution relates to this swapping procedure. The D-H algorithm leverages an essential feature of SPVs—each SPV has its own destination (and latest arrival time), which heavily constrains the possible routes for each SPV. As such, we can enumerate all candidate routes for each SPV. Then, when assigning PDOs to SPVs, we assign PDOs to specific SPV routes. The computational experiments illustrate that the proposed D-H is much faster and more scalable than an exact solution approach. For small problem instances, the optimality gap between the D-H and the exact solution is quite small (i.e., a 0.5-2.7% optimality gap).

3. The third contribution lies in the D-H algorithm's swapping procedure. Instead of applying a standard or conventional heuristic or meta-heuristic for a VRP, we tailor a swapping procedure that combines "local search" with "relatively larger neighborhood search" in an SA-inspired procedure. The proposed swapping procedure effectively improves the solution quality without requiring significant computational effort. We compare our swapping procedure with a large neighborhood search (LNS) heuristic and find that our proposed swapping procedure is competitive with the LNS algorithm regarding solution quality, but the swapping procedure is substantially faster than the LNS algorithm.
4. The study also contributes to industrial practice by analyzing the impact of critical CSD system design factors on key logistics performance metrics. For example, we analyze the impact of (i) additional SPVs on total delivery costs, (ii) CD driver willingness to detour on total vehicle miles and carbon emissions, and (iii) CD drivers rejecting PDOs on total delivery costs.

The remainder of this paper is organized as follows. Section 2 reviews the literature relevant to CD. Section 3 formulates the underlying problem using a conventional VRP-based model and a set partitioning model. Section 4 describes the novel D-H solution algorithm. Section 5 introduces a real-world numerical case study in Irvine, CA, USA. Section 6 includes the computational results. Finally, Section 7 summarizes major findings, concludes the paper, and discusses future research.

# 2 Literature Review

This section reviews literature relevant to the proposed CSD service defined in Section 1. Section 2.1 provides general background information on CD for the last-mile logistics problem. Section 2.2 focuses on studies that model and solve CD operational problems.

## 2.1 Crowdsourced Last-mile Delivery

We distinguish between two types of studies related to CD. The first type includes empirical studies that rely on collecting behavioral data, conducting surveys, or performing experiments. The studies in this category attempt to answer behavioral questions related to CD. Examples include understanding customer acceptance of CD (Punel et al., 2018; Punel and Stathopoulos, 2017) and the willingness of drivers to participate in CD (Castillo et al., 2022; Ermagun et al., 2020; Ermagun and Stathopoulos, 2018; Miller et al., 2017). These studies provide qualitative and quantitative insights into the behavioral aspects of CD. These behavioral insights also provide value to the second type of CD study.

The second type of CD studies uses mathematical models (e.g., analytical models, optimization models, simulations) to optimize or analyze specific CD services. Recent studies in this area focus on (i) workforce planning, acquisition, and scheduling, (ii) pricing and compensation schemes, (iii) delivery task assignment and vehicle routing, and (iv) integration of CD with other services. We provide more details on each of these four areas below.

Studies related to workforce planning focus on the supply of CD workers. Some studies focus on static capacity planning with a mix of dedicated and occasional drivers (Dai and Liu, 2020; Tao et al., 2021). Others focus on planning with a mix of (i) contract drivers who commit to a specific period and (ii) occasional drivers (Yildiz and Savelsbergh, 2019). Apart from the problem of static capacity planning, problems of dynamic workforce acquisition and scheduling are also studied under this category (Behrendt et al., 2023; Lei et al., 2020; M. Ulmer and Savelsbergh, 2020). Depending on the communication method between CD drivers and the platform, the supply workforce acquisition is sometimes achieved through task menu selection by drivers (Ausseil et al., 2022).

Studies related to compensation schemes for CD drivers focus on pricing mechanisms to attract drivers while minimizing driver acquisition costs. Qi et al. (2018) apply an analytical approach to determine the minimum compensation for driver acquisition. M. Ulmer (2020) studies the problem of dynamic pricing in the same-day delivery problem. Barbosa et al. (2023) apply a data-driven approach to consider the scenarios where drivers may reject PDO matches due to the low compensation offered. Notably, workforce planning is directly related to driver compensation since pricing and compensation schemes impact drivers' willingness to participate.

Most CD studies, including the current study, focus on delivery task assignment and vehicle routing (Savelsbergh and Ulmer, 2022). Section 2.2 reviews this area in detail.

In the last area of studies, prior research integrates CD service with other services. One extension of the fixed depot CD service problem involves deploying flexible or mini depots in the service design. The CD problem with flexible depots permits the transshipment of parcels between CD drivers. Mousavi et al. (2022) assume stochastic demand for PDOs and integrate the last-mile CD with mobile depots. Nieto-Isaza et al. (2022) deploy multiple mini-depots to reduce transportation costs while minimizing the depot deployment cost. Like mini depots, researchers also integrate parcel lockers/transshipment points with CD service for last-mile delivery. Parcel lockers are different from mini depots as the former permit customers to pick up PDOs at the lockers (Ghaderi et al., 2022; Yu et al., 2022). Another application of CD service is the integration with item-sharing (Behrend and Meisel, 2018). Finally, designs of the last-mile CD service also involve the participation of neighbors (Akeb et al., 2018) or friends (Devari et al., 2017), in which drivers can deliver parcels to neighbors when the final PDO recipients are not at home.

## 2.2 Delivery Task Assignment and Vehicle Routing in Crowdsourced Delivery

Delivery task assignment and vehicle routing are fundamental components of CD services. As such, most studies that aim to quantitatively analyze the supply-side of CD service rely directly or indirectly on models and algorithms for delivery task assignment and routing decisions. In this section, we distinguish between dynamic and static as well as stochastic and deterministic delivery task assignment and routing decision problems.

Dynamic CD problems involve making sequential decisions for task assignment and routing, where information about future delivery requests is incomplete. Arslan et al. (2019) propose a seminal model for dynamic CD. In addition to the case of a single store as the depot, they also consider in-store customers' willingness to travel to another depot for pickup and delivery. To solve the problem, they propose a rolling horizon approach that employs a matching problem solution technique. Dayarian and Savelsbergh (2020) model the dynamic CD problem, assuming a maximum of one task per SPV. They propose and compare decision strategies, namely, myopic assignment and sample scenario planning. Archetti et al. (2021) model the dynamic problem where CD drivers can simultaneously serve multiple PDOs. They assume stochasticity associated with PDOs, but they assume complete information on CD drivers, including a pre-specified working time window. Fatehi and Wagner (2022) apply an analytical approach to dynamically match tasks to either CD drivers or 3PL logistic providers (dedicated drivers), but they do not try to optimize vehicle routes.

Other than problems that consider task assignment and routing, Zehtabian et al. (2022) propose a dynamic model to estimate the arrival time of pickup and delivery of CD drivers and solve the problem with a look-ahead policy. To handle larger-scale problems, Simoni and Winkenbach (2023) propose a dynamic task assignment with PDO batching to improve assignment efficiency.

Stochastic problem settings include uncertainty from the supply side (e.g., stochastic driver supply, driver information, driver rejection), the demand side (e.g., uncertain delivery requests, uncertain customer availability/presence), or both. Most stochastic problems deal with supply-side uncertainty (Pugliese et al., 2023; Feng et al., 2021; Gdowska et al., 2018; Torres et al., 2022a, 2022b). Gdowska et al. (2018) consider the case where in-store customers (i.e., CD drivers) may reject offers from the central manager; they model the problem as a bi-level stochastic program. They solve the bi-level stochastic program with a cost-driven heuristic technique. Both Feng et al. (2021) and Torres et al. (2022a) consider the case where the destinations of CD drivers are uncertain, while Torres et al. (2022b) assume that CD driver supply and available time windows are unknown. They apply a branch and price algorithm to solve small instances and a heuristic for large instances. Pugliese et al. (2023) capture travel time uncertainty and incorporate penalties for arriving after a time window; they solve the stochastic model with Bender's decomposition and column generation. Finally, Silva et al. (2023) consider stochasticity in driver supply and customer demand, where there is a correlation between the two sources of uncertainty. They solve the stochastic model with a branch-price-and-cut algorithm.

Although not in the context of CD, studies on last-mile e-commerce delivery consider uncertainty in terms of the presence/availability of customers at delivery locations (Özarık et al., 2021, 2023). However, these two studies focus on "attended home delivery" with "valuable, perishable, or oversized items" (Özarık et al., 2021, 2023), whereas, in this study, we focus on standard, non-perishable, non-luxury goods and assume that drivers can drop off packages without interacting with customers.

Deterministic problems assume complete information about PDOs and CD drivers. Studies in the literature usually formulate the deterministic CD problem as a multi-vehicle routing problem (m-VRP) or multi-vehicle pickup and delivery problem (m-PDP).

Some studies only consider CD drivers and omit DVs, e.g., Wang et al. (2016), Behrend and Meisel (2018), Mofidi and Pazour (2019), and Le et al. (2021). Wang et al. (2016) model the problem as a minimum cost flow problem and solve relatively large-scale problem instances. Behrend and Meisel (2018) integrate item-sharing with CD and solve the item-sharing and crowd-shipping problem using the exact method proposed by Behrend et al. (2019). Mofidi and Pazour (2019) apply a two-stage bilevel program for the hierarchical decision of the CD problem, where the upper-level decision is the assignment of PDOs to drivers, and the lower-level decision is the driver's choice of specific PDOs. Le et al. (2021) assume a similar "shared trip" setting for CD drivers as this paper. However, they focus on comparing compensation schemes that consider the shipper's minimum willingness to pay and drivers' minimum expected compensation. Another study by Boysen et al. (2022) considers store employees as the workforce to deliver PDOs. However, none of the studies in this paragraph consider the routing of DVs.

Ghaderi, Tsai, et al. (2022) propose a two-step task assignment and vehicle routing model. In the first step, they profile the historical trajectory of CD drivers to obtain driver routes. In the second step, they match delivery tasks to historical CD driver routes. However, by only using historical trajectories, their method is unlikely to find optimal matches and vehicle routes.

Several studies in the literature focus on the deterministic problem and consider assignment and routing decisions for both CD and DV drivers. Table 1 compares the studies most relevant to ours. Archetti et al. (2016) formulate the problem as a classic VRP. Their model assumes a maximum of one task per SPV driver, and they apply a multi-start heuristic by first assigning all PDOs to dedicated trucks and then solving a series of small-scale bi-partite matching problems to assign PDOs to SPVs. Macrina et al. (2017) extend the problem to a VRP with time windows (VRPTW) and allow SPVs to carry multiple PDOs. Dahle et al. (2019) consider pickups and drop-offs and formulate the problem as a pickup and delivery problem with time windows (PDPTW). Dahle et al. (2019) focus on comparing different compensation schemes; they argue that compensation needs to be large enough to exceed the threshold of the driver's willingness to deliver. Dahle et al. (2019) conclude that CD can reduce total costs for logistic companies by 10-15%.

Unlike the prior three studies in Table 1, Kafle et al. (2017) and Macrina et al. (2020) consider transshipment in CD. Kafle et al. (2017) assume a bidding procedure for CD drivers before assigning PDOs. They formulate the problem as a generalized VRPTW. They solve the master problem by decomposing it into two subproblems, namely, bidding selection and routing/scheduling of trucks. Macrina et al. (2020) formulate the CD problem with transshipment nodes as a generalized version of the VRP with occasional drivers. They propose a variable neighbor descent heuristic to solve the problem.

Another study that is worth reviewing but does not appear in the table is Ahamed et al. (2021). They also route CD drivers and DVs. However, the ad-hoc CD drivers in their study are more similar to committed drivers than CSD drivers, and their compensation depends on total distance instead of detour. Ahamed et al. (2021) apply a deep reinforcement learning approach to solve the deterministic problem with instances as large as 200 PDOs and 70 drivers.

Formulating the CD problem based on the VRP or PDP is quite natural. However, due to the NP-hard nature of routing problems, and despite the rich literature on VRPs (Cordeau and Laporte, 2003; Golden et al., 2008; Laporte, 1992; Laporte et al., 2000), obtaining optimal routes is difficult when the number of DVs exceeds just 10-15, and there are no SPVs. Problems with hundreds of potential SPVs, in addition to DVs, such as the one addressed in this study, would be near-impossible to solve.

**Table 1: Comparison of Related Literature**

| Literature | Formulate | SPV Capacity | SPV Compensation | DV Constraint | Solution Technique | Test Scale |
|---|---|---|---|---|---|---|
| Archetti et al. (2016) | VRPOD | One | Package Location | Capacity | Multi-Start Heuristic | 100 Tasks, 100 SPVs |
| Macrina et al. (2017) | VRPOD-based | Multiple | Detour | Capacity | CPLEX | 100 Tasks, 100 SPVs |
| Dahle et al. (2019) | PDPTW-based | Multiple | Thresholds | Capacity | MOSEL | 70 Tasks, 100 SPVs |
| Kafle et al. (2017) | VRPTW-based | Multiple | Bidding | Capacity | Decompose into Two Subproblems | 100 Tasks, 60 SPVs |
| Macrina et al. (2020) | VRPOD-based | Multiple | Detour | Capacity | Variable Neighbor Descent | 100 Tasks |
| ***This Paper*** | **Set Cover-Based** | **Max. 4** | **Fixed + Detour** | **Capacity + Max Hours** | **Decomposition Heuristic** | **900 Tasks, 1200 SPVs** |

Table 1 provides a comparison of studies addressing the deterministic CD problem. Notably, the current paper differentiates itself from closely related studies along several dimensions. First, we extend the original VRP with occasional drivers to a more generalized form that applies to DV and SPVs. The SPV drivers' origins are also generalized instead of being only at the depot or in-store. We model the generalized problem using a traditional VRP-based formulation and an alternative set-partitioning formulation. Second, we employ a scheme compensating SPV drivers per PDO and detour mile. Third, the scale of the problem instances we address is much larger than in previous studies, as highlighted in the last column of Table 1. Fourth, we propose a novel decomposition heuristic that we tailored explicitly for large instances of the CSD PDO-vehicle matching and vehicle routing problem with SPVs and DVs.

# 3 Problem Description and Mathematical Formulations

**Table 2: Notations and Acronyms**

| Notation | Description |
|---|---|
| $A$ | The arc set |
| $a_{prs}$ | Binary parameter, whether a PDO $p$ could be served by $r^{th}$route of SPV $s$ |
| $a_{ijs}$ | Binary parameter, whether a node $j$ could be visited by $i^{th}$route of SPV $s$ |
| $a_{ijd}$ | Binary parameter, whether a node $j$ could be visited by $i^{th}$route of DV $d$ |
| $B$ | Travel budget (in minutes) |
| $B_s$ | Travel budget of SPV $s$ (in minutes) |
| $B_s^M$ | The maximum willingness of SPV $s$ to detour (in minutes) |
| $c_{ij}^s$ | Monetized travel cost to use link $(i,\ j)$ for SPV |
| $c_{ij}^d$ | Monetized travel cost to use link $(i,\ j)$ for DV |
| $c_s^o$ | Monetized cost from origin to hub 0 for SPV $s$ |
| $c_s$ | Monetized cost from origin to destination for SPV $s$ |
| $c_{is}$ | Cost of using route $i$ for SPV $s$ |
| $c_{id}$ | Cost of using route $i$ for DV $d$ |
| $c_{ir}$ | Cost of using route $r$ to serve PDO $i$ |
| $c_{rs}$ | Cost of route $r$ for SPV $s$ |
| $D$ | Set of dedicated vehicles (DVs) |
| $d$ | Individual DV, $d \in D$ |
| $e^d$ | Coefficient for distance-based compensation for SPs |
| $e^o$ | Coefficient for per-PDO compensation for SPVs |
| $F_d$ | Fixed cost of using DV $d$ |
| $G = (N, A)$ | Network $G$ consists of Vertex/Node Set $N$ and Arc/Link Set $A$ |
| $(i, j)$ | A tuple to describe a link between node $i$ and node $j$ |
| $h$ | Distribution center node where DVs return after completing a route |
| $k$ | An individual vehicle, either a DV or a SPV |
| $\lambda_i$ | Decision variable in the dual problem corresponding to constraint $i$ |
| $M$ | A large number |
| $N$ | Node set |
| $n_p$ | Drop-off node for PDO $p$ |
| $n_s$ | Destination node for SPV $s$ |
| $o$ | Depot |
| $P$ | Set of PDOs |
| $p$ | Individual PDO, $p \in P$ |
| $q_d$ | The maximum number of stops for a DV route |
| $q_s$ | The maximum number of PDOs that an SPV $s$ is willing to serve |
| $R$ | The set of feasible routes across all SPVs |
| $R_{reduce}$ | The set of feasible routes for only the SPVs assigned to PDOs during initial matching (i.e. in Step 2) |
| $S$ | Shared-personal vehicle (SPV) set |
| $s$ | Individual SPV, $s \in S$ |
| $\tau_{ij}$ | Time cost to use link $(i, j)$ |
| $\tau_{pd}$ | PDO pickup delay, constant |
| $T^D$ | The maximum working time of DV |
| $T_d^s$ | The earliest time that SPV $s$ could depart from its origin depot $o$ |
| $T_a^s$ | The latest time that SPV $s$ should arrive at its destination depot $h$ |
| $T_d^p$ | Earliest pickup time for PDO $p$ |
| $T_a^p$ | Latest arrival time for PDO $p$ |
| $t_i^k$ | The time vehicle $k$ arrives at node $i$ |
| $\Theta$ | Objective function value |
| $u_d$ | Binary variable indicating whether the optimal solution includes DV |
| $V$ | Total vehicle set |
| $\omega$ | Service reward adjustment factor, a relatively large number |
| $x_{ij}^k$ | Binary variable indicating whether vehicle $k$ traverses link $(i, j)$ |
| $x_{prs}$ | Binary variable indicating whether SPV $s$ serves PDO $p$ on route $r$ |
| $y_{is}$ | Binary variable indicating whether SPV $s$ uses its $i^{th}$ feasible route |
| $y_{id}$ | Binary variable indicating whether DV $d$ uses its $i^{th}$ feasible route |
| $z_{rs}$ | Binary variable indicating whether SPV $s$ uses route $r$ |
| $z_s$ | Binary variable indicating whether SPV $s$ carries any PDOs |

| Acronym | Description |
|---|---|
| CD | Crowdsourced Delivery (specifically for the last mile in this paper) |
| CSD | Crowdsourced Shared-trip Delivery |
| D-H | Decomposition Heuristic |
| DV | Dedicated Vehicle |
| LNS | Large Neighborhood Search |
| MFOCVRPTW | Mix Fleet Open Capacitated Vehicle Routing Problem with Time Window |
| PDO | Package Delivery Order |
| PDP(TW) | Pickup and Delivery Problem (with Time Window) |
| SA | Simulated Annealing |
| SPV | Shared Personal Vehicle |
| VRP(TW) | Vehicle Routing Problem (with Time Window) |
| VRPOD | Vehicle Routing Problem with Occasional Drivers |

### 3.1 Problem Description

Let $P$ denote the set of PDOs that require delivery, indexed by $p \in P$ . Each PDO $(p)$ may include multiple items. We assume that all items are small- to medium-sized and easily fit in a standard sedan. We also assume that the packages are standard (non-luxury) non-perishable goods and do not require the customer to receive the package. These assumptions are consistent with existing CD services, including Amazon Flex. Each PDO $p$ has a designated drop-off location, an earliest pickup time, $T_d^p$, and latest delivery time, $T_a^p$.

Two types of vehicles are used for delivery in the CSD system: SPVs (usually family-size sedans, wagons, mini-vans, or SUVs) and DVs (usually medium-duty vans or trucks) for delivery. Let $V$ be the set of all vehicles, $S$ be the set of SPVs, and $D$ be the set of DVs; hence, $V = \{S \cup D\}$. An individual SPV is represented as $s$ $(s \in S)$. The driver of an SPV indicates the maximum number of PDOs they are capable or willing to carry/serve, and the parameter is denoted $q_s$. Each SPV has its own origin and destination pair. If any PDOs are assigned to an SPV, the SPV must first travel from its origin to the depot to pick up the PDO(s); second, the SPV must deliver all PDO(s); and lastly, the SPV must travel to its own destination. Let $T_d^s$ denote the earliest time an SPV $s$ can pick up PDOs at the depot and let $T_a^s$ denote the latest arrival time that an SPV $s$ should arrive at its own destination.

Let $d$ denote a DV $(d \in D)$. In addition to $d$ indexing the DV set $D$, and $s$ indexing the SPV set $S$, let $k$ index the set of all vehicles $V$ . Without loss of generality, we assume all DVs are identical and have a maximum operating time, denoted as $T^D$. We also assume each DV has a maximum number of PDOs that it can carry, denoted as $q_d$. DVs are required to return to the depot/hub after completing delivery tasks.

The service network is defined on a graph $G = (N, A)$. $N$ is the set of nodes, including the hub, all PDO drop-off locations, and all origins and destinations of SPVs. $A$ is the set of arcs/links connecting nodes, represented by the tuple $(i, j)$, where $i, j$ are nodes. Let $o \in N$ denote the depot where DVs depart and let $h \in N$ denote the depot where DVs terminate their routes. Physically, $o$ and $h$ are both the depot. The drop-off location of a PDO $p$ is represented as $n_p$. The designated destination node of an SPV $s$ is represented as $n_s$.

The monetized travel cost of a link $(i, j)$ is represented as $c_{ij}^s$ and $c_{ij}^d$ for SPVs and DVs, respectively. The travel time of a link $(i, j)$ is represented as $\tau_{ij}$. An SPV driver is compensated by the number of PDOs completed and the total detour distance from delivering PDOs. The compensation per PDO is represented by $e$. The detour distance calculation is demonstrated in Figure 2. The total detour cost (compensation) for an SPV is calculated as $c_s^o + c_d - c_s$, where $c_s^o$ is the monetized cost for an SPV to travel from its origin to the depot; $c_s$ is the monetized cost for each SPV to travel directly from its origin to its destination; and $c_d$ is the delivery route distance from the depot to the SPV's destination.

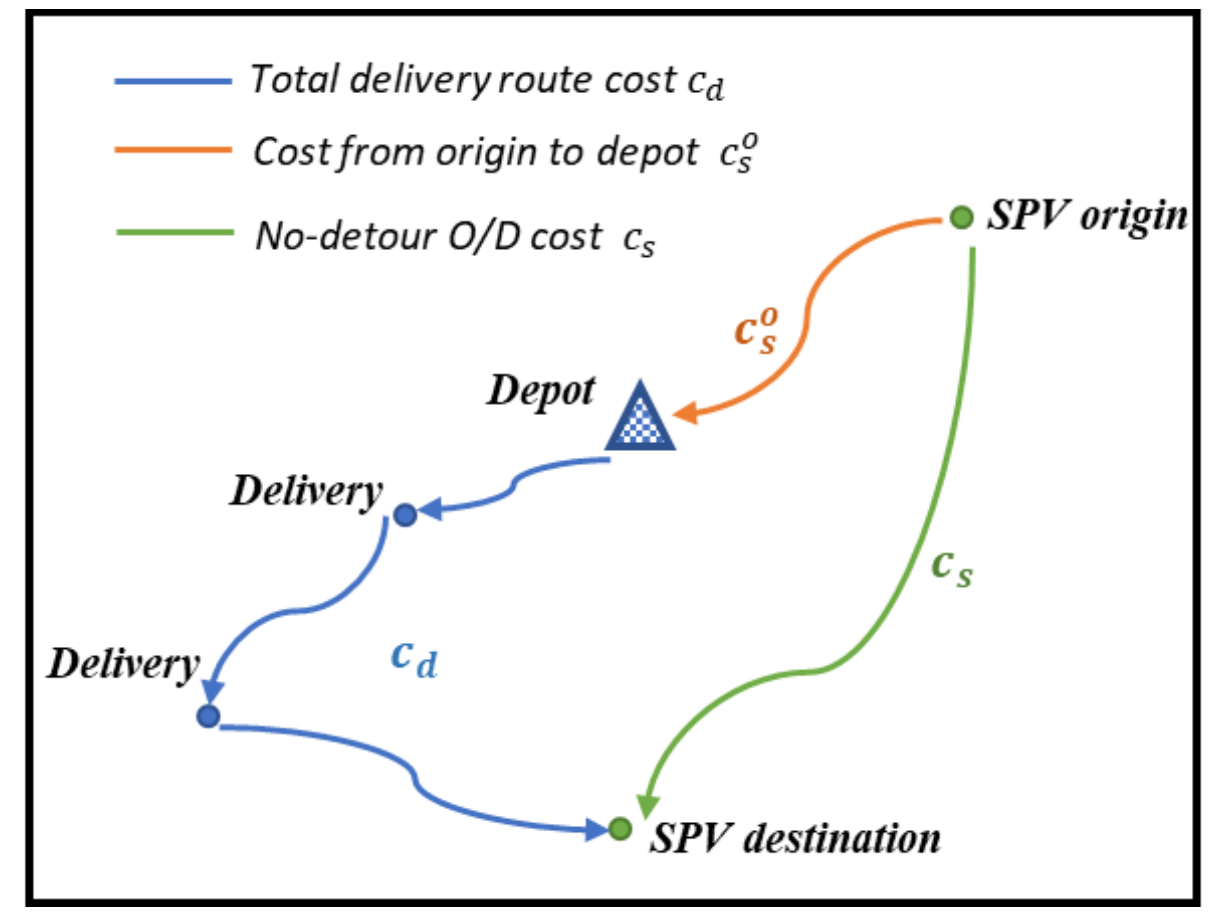

**Figure 2: Method for calculating SPV detour distance**

## 3.2 m-VRP-based Formulation

The natural way to formulate a delivery problem involves exploiting VRP-based formulations since the delivery problem has the features of an “unrepeated route” and “returning to the depot.” The CD problem requires a generalization of the original VRP. First, we must route two vehicle types, SPVs and DVs. Therefore, the problem is a heterogeneous- or mixed-fleet VRP (Baldacci et al., 2008; Irnich et al., 2014). In addition, in this CSD problem, SPVs do not return to the depot, which makes the problem similar to another variant of VRP, the so-called Open VRP (Li et al., 2007). Hence, we classify the CSD problem with SPVs and DVs as a *Mixed Fleet Open Capacitated Vehicle Routing Problem with Time Windows (MFOCVRPTW)*.

The decision variables are as follows:

- $z_s \in \{0,1\}, \forall\, k \in S.$ — $z_s = 1$, if SPV $s$ is used.
- $u_d \in \{0,1\}, \forall k \in D.$ — $u_d = 1$, if DV $d$ is used.
- $x_{ij}^k \in \{0,1\}, \forall (i,j) \in A, \forall\, k \in V.$ — $x_{ij}^k = 1$, if arc $(i,j)$ is visited by vehicle $k$.
- $t_i^k \in \mathbb{R}^+, \forall i \in N, \forall k \in V.$ — Arrival time of vehicle $k$ at node $i$.

The formulation of the MFOCVRPTW is as follows.

### 3.2.1 Formulation 1

$$\min_{x,z,t,u} \boldsymbol{\Theta_1} = \sum_{k,s \in S} \left\{ z_s \left[ e^d \left( c_s^o + \sum_{(i,j)\in A} c_{ij}^s x_{ij}^k - c_s \right) + e^o \left( \sum_{(i,j)\in A} x_{ij}^k - 1 \right) \right] \right\} + \sum_{k \in D} \sum_{(i,j)\in A} c_{ij}^d x_{ij}^k + \sum_{d \in D} F_d u_d \tag{1}$$

subject to

$$\sum_{j \in \{n_p, n_s\}, j \neq i} \sum_{k \in V} x_{ij}^k = 1 \qquad \forall i \in \{n_p\} \tag{2}$$

$$\sum_{i\in\{o,n_p\},i\neq j}\sum_{k\in V} x_{ij}^k = 1 \qquad \forall j \in N\setminus\{h\} \qquad (3)$$

$$\sum_{i\in\{o,n_p\},i\neq j}\sum_{k\in V} x_{ij}^k - \sum_{l\in\{n_p,n_s\},l\neq j}\sum_{k\in V} x_{jl}^k = 0 \qquad \forall j \in \{n_p\} \qquad (4)$$

$$\sum_{i\in N\setminus\{n_s\}} x_{i,n_s}^k = 1 \qquad \forall k \in S \qquad (5)$$

$$\sum_{j\in\{n_p\}} x_{oj}^k - \sum_{i\in\{n_p\}} x_{i,h}^k = 0 \qquad \forall k \in D \qquad (6)$$

$$u_d \geq \sum_{j\in\{n_p\}} x_{oj}^k \qquad \forall d,k \in D \qquad (7)$$

$$\sum_{(i,j)\in A\setminus\{(0,n_s)\}} x_{ij}^k \leq z_s \times (q_s + 1) \qquad \forall s,k \in S \qquad (8)$$

$$\sum_{(i,j)\in A} x_{ij}^k \leq u_d \times \left(q_d + 1\right) \qquad \forall d,k \in D \qquad (9)$$

$$T_d^s \leq t_o^k \qquad \forall s,k \in S \qquad (10)$$

$$t_{N_s}^k \leq T_a^s \qquad \forall s,k \in S \qquad (11)$$

$$t_h^k \leq T^D \qquad \forall k \in D \qquad (12)$$

$$T_d^p \leq t_o^k + \left(1 - \sum_{i\in N\setminus\{n_s\}} x_{i,n_p}^k\right) \times M \qquad \forall p \in P, \ \forall k \in S \qquad (13)$$

$$t_{n_p}^k \leq T_a^p + \left(1 - \sum_{i\in N\setminus\{n_s\}} x_{i,n_p}^k\right) \times M \qquad \forall p \in P, \ \forall k \in S \qquad (14)$$

$$t_i^k + \tau_{ij} \leq t_j^k + \left(1 - x_{ij}^k\right) \times M \qquad \forall i,j \in N,\ i \neq j \ \ \forall k \in V \qquad (15)$$

$$x_{ij}^k \in \{0,1\} \qquad \forall i,j \in N \ \ \forall k \in V \qquad (16)$$

$$z_s \in \{0,1\} \qquad \forall s \in S \qquad (17)$$

$$u_d \in \{0,1\} \qquad \forall d \in D \qquad (18)$$

$$t_i^k \geq 0 \qquad \forall i \in N, \forall k \in V \qquad (19)$$

In this formulation, the objective function aims to minimize the total cost of delivery using SPVs and DVs. The first (long) term is the total SPV cost, which consists of distance- and per-PDO-based costs. The term $e^d\left(c_s^o + \sum_{(i,j)\in A} c_{ij}^s x_{ij}^k - c_s\right)$ represents the total distance (detour) cost of an SPV $k$. The term $e^o\left(\sum_{(i,j)\in A} x_{ij}^k - 1\right)$ is the compensation paid to an SPV based on the number of fulfilled PDOs. $e^d$ and $e^o$ are two parameters (coefficients), which are the weights of distance-based and per-PDO-based rewards in the total compensation. The relative magnitude of $e^d$ and $e^o$ determine the cost structure. For example, when $e^d = 0$, drivers are rewarded only based on the number of PDOs fulfilled and not detour distance, and vice versa for $e^o = 0$.

The first term is multiplied by $z_s$, indicating whether SPV $s$ is used. If $z_s$ is 0, the SPV travels directly from its origin to its destination. The second term, $\sum_{k \in D} \sum_{(i,j) \in A} c_{ij}^d x_{ij}^k$, is the total delivery routing cost of all DVs. The last term, $\sum_{d \in D} F_d u_d$, calculates the total fixed cost for using DVs.

(Additionally, we want to note that the distance-based cost term is quite general. For example, if we want to compensate drivers based on total distance rather than detour distance, we can set $c_s$ to zero.)

Constraints (2) to (6) are routing constraints. Constraints (2) and (3) indicate that every node must be visited once and only once by a vehicle (of either type). The constraints in Eqn. (4) are the flow balance constraints at each node. The constraints in Eqn. (5) indicate that each SPV must arrive at its designated destination. The constraints in Eqn. (6) require that if a DV leaves the depot, it must return to the depot. The constraints in Eqn. (7) ensure that only "activated" DVs can serve requests. The constraints in Eqn. (8), together with the constraints in Eqn. (5), ensure that when $z_s = 0$, SPV $s$ travel directly to its own destination ($n_s$), and does not serve PDOs. Additionally, the constraints in Eqn. (8) regulate the number of delivery locations that an SPV $s$ can visit and is the so-called "capacity" constraint. Similarly, the constraints in Eqn. (9) regulate the number of delivery locations a DV can visit. In addition to Eqn. (9), DVs are also constrained by the total number of working hours/distances, represented by Eqn. (12).

Constraints (10) to (15) are time window constraints. The constraints in Eqn. (10) guarantee that the trip for an SPV starts after its earliest departure time. The constraints in Eqn. (11) guarantee that an SPV's arrival time at its destination must be no later than its latest arrival time. The constraints in Eqn. (13) ensure that a PDO is only picked up after its earliest pickup time. The constraints in Eqn. (14) ensure that a PDO must be delivered no later than its latest delivery time. The constraints in Eqn. (15) indicate that if $x_{ij}^k = 1$ (i.e., node $j$ is visited right after node $i$ by vehicle $k$), the arrival time of vehicle $k$ at node $j$ must be later than the arrival time of vehicle $k$ at node $i$ plus the necessary travel time of arc $(i, j)$. The constraints in Eqn. (15) also serve as sub-tour elimination constraints. The constraints in Eqn. (16) to (19) are the binary and non-negativity constraints for decision variables.

Formulating the CSD problem from a vehicle routing perspective enables us to solve the problem by leveraging the rich literature on VRP. Exact methods include Branch-and-bound (Christofides and Eilon, 1969; Little et al., 1963) and branch-and-cut or generating cuts (Baldacci et al., 2008; Baldacci et al., 2012; Laporte et al., 1985). Heuristics include the Clarke and Wright saving heuristic (Clarke and Wright, 1964) and multiple meta-heuristics (Gendreau and Potvin, 2005; Hansen et al., 2001; Nikolaev and Jacobson, 2010; Prins, 2004).

The *MFOCVRPTW* is an NP-hard problem, and using exact methods for large-scale problem instances is computationally infeasible. Therefore, heuristics are preferable for large-scale problems. We construct a decomposition-based heuristic (D-H) algorithm to solve the problem. The basis for the decomposition is a set partitioning formulation, presented in the following subsection.

### 3.3 Set Partitioning Formulation

The *m-VRP* can be reformulated as a set partitioning problem (Baldacci et al., 2008; Desrosiers et al., 1992; Laporte, 1992; Y. H. Lee et al., 2008; Ropke and Cordeau, 2009). This paper treats the collection of all PDO locations $\{n_p\}$ as a set of nodes to be covered/contained by a collection of route sets (the collections of vehicle routes). Then, the problem is to assign each PDO delivery location to one feasible vehicle route while minimizing the total cost of the collection of vehicle routes.

Like the approaches of Baldacci et al. (2008) and Ropke and Cordeau (2009), we formulate the CSD problem as a set partitioning problem. Let $y_{is}$ be a binary decision variable and represent whether the $i^{th}$ feasible route of SPV$s$ is used. Correspondingly, let the binary variable $y_{id}$ represent whether the $i^{th}$ feasible route of DV $d$ has been used. Let $c_{is}$ and $c_{id}$ be the cost of traveling on route $i$ for SPV $s$ and DV $d$, respectively. Let $a_{ijs}$ and $a_{ijd}$ be two binary parameters equal to 1 if it is feasible for $route\ i$ of shared $(s)$ or DV $(d)$ to service PDO $j$. Formulation 2 provides the set partitioning formulation of the MFOCVRPTW.

#### 3.3.1 Formulation 2

$$\min_{y} \ \mathbf{\Theta_2} = \quad e^d \sum_s \sum_i (c_{is} - c_s)\, y_{is} + e^o \sum_s \sum_i \sum_j a_{ijs} y_{is} + \sum_d \sum_i c_{id} y_{id} + F_d \sum_d \sum_i y_{id} \qquad (20)$$

subject to

$$\sum_s \sum_j a_{ijs} y_{is} + \sum_d \sum_j a_{ijd} y_{id} = 1 \qquad \forall i \in \{n_p\} \qquad (21)$$

$$\sum_i y_{is} = 1 \qquad \forall s \in S \qquad (22)$$

$$\sum_i y_{id} \leq 1 \qquad \forall d \in D \qquad (23)$$

$$\sum_i \sum_j a_{ijs} y_{is} \leq q_s \qquad \forall s \in S \qquad (24)$$

$$\sum_i \sum_j a_{ijd} y_{id} \leq q_d \qquad \forall d \in D \qquad (25)$$

$$y_{is}, y_{id} \in \{0,1\} \qquad \forall (i,s) \in R, \forall (i,d) \in R \qquad (26)$$

The objective function in Eqn. (20) is similar to Eqn. (1) and minimizes the total cost. The first term, $e^d \sum_s \sum_i (c_{is} - c_s)\, y_{is}$ is the total detour cost of SPVs, and the second term $(e^o \sum_s \sum_i \sum_j a_{ijs} y_{is})$ is the total compensation per PDO fulfilled for SPVs. The third term is the total routing cost of DVs, and the last term is the total fixed cost of using DVs.

The constraints in Eqn. (21) ensure that each PDO must appear once and only once on all vehicle routes. The constraints in Eqn. (22) guarantee that one and only one feasible route for each SPV is selected in the optimal solution. The constraints in Eqn. (23) state that no more than one feasible route for a DV should be used. The constraints in Eqn. (24) and (25) limit the number of stops SPVs and DVs can make, respectively.

It is possible to build these two constraints into the construction of vehicle routes. The constraints in Eqn. (26) are binary constraints for the decision variables.

A set partitioning problem can easily be converted to a bi-partite matching problem. Therefore, Formulation 2 provides a new starting point for solving the CSD problem as a matching/assignment problem between PDOs and vehicle routes. Since the bi-partite matching problem has the feature of total unimodularity (Yannakakis, 1985), it allows a linear relaxation of an integer problem.

However, challenges still remain. It is necessary to enumerate all possible routes for each SPV and DV to ensure optimality. Generating all feasible routes for SPVs is possible because of the tight time-window constraints on SPV trips/routes. The ability to enumerate all possible SPV routes makes the set partitioning formulation quite appealing for the CSD problem. However, generating all possible DV routes is computationally infeasible. Hence, most research that employs set partitioning to handle the standard VRP problem relies on generating a 'sufficient' number of promising routes for each vehicle (Ryan et al., 1993).

Given that it is feasible to generate all SPV routes but not feasible to do the same for DVs, we propose decomposing the problem by vehicle type. Hence, in the next section, we propose a Decomposition-based Heuristic (D-H) to solve the CSD problem.

# 4 Decomposition Heuristic

## 4.1 Solution Algorithm Overview

The four major D-H algorithmic steps include SPV route generation, PDO-SPV assignment, DV routing, and PDO switching between DVs and SPVs, summarized in Steps 1 through 4 below. The D-H algorithm has two major components: obtaining a quality initial solution (Steps 1, 2, and 3) and an iterative procedure to improve the solution (Step 4). Like Arslan et al. (2019), we also assume that the average cost of using SPVs for PDO delivery is lower than that of using DVs.

- ***Step 1****: SPV route generation*
    - *Generate the set of feasible routes for each SPV in $S_0$.*
        - *Most of these routes can be generated offline ← described below.*
- ***Step 2:*** *PDO-SPV route assignment*
    - *Assign PDOs to SPVs. This problem is similar to a bi-partite matching problem and can be efficiently solved for large problem instances. The primary objective of the matching problem is to maximize PDOs assigned to SPVs.*
    - *For each SPV assigned to a PDO, store these SPVs and their feasible routes for consideration in Step 4.*
- ***Step 3:*** *DV routing*
    - *Run a multi-vehicle routing problem insertion heuristic to assign unassigned (after Step 2) PDOs to DVs and route DVs.*
    - *At the end of Step 3, we have an initial feasible solution and associated cost.*
- ***Step 4:*** *Swapping PDOs between SPVs and DVs for solution improvements*
    - *Run SA-inspired swapping procedure to exchange PDOs between SPVs and DVs iteratively.*
        - *The procedure and rules of switching are described in Section 4.4*

The next few subsections describe these four steps in more detail.

### 4.2 Step 1: Shared Vehicle Route Generation with Budgeted K-shortest Paths

This section describes the first step of the D-H, where we generate the k-shortest paths subject to budget constraints. We generate the possible routes from the depot to different SPV destination locations for various values for willingness to detour. The quality of the *k-shortest paths* is a primary determinant of the solution quality of the PDO-SPV route assignment problem, and we attempt to exhaustively generate all routes for an SPV under a travel budget constraint.

Unlike the other steps in the D-H, the logistics platform can compute and store a large combination of SPV routes once (e.g., at the beginning of the month or year), as opposed to needing to recompute the routes every time (e.g., every morning) they solve a CSD problem instance. Hence, we say that this subproblem can be addressed "offline." It is possible to compute and store these SPV routes because (i) there are a limited number of delivery locations in the region, (ii) there are a limited number of SPV trip origins and destinations in the region, (iii) the maximum detour for any SPV is quite limited, and (iv) an SPV can only serve a small number of PDO locations in one route. Then, when solving a CSD problem instance, we can "activate" the set of SPV routes that are consistent with the SPV drivers for a given problem instance. Handling Step 1 offline is beneficial because it is computationally intensive, and therefore, the logistics provider can expend computation resources on Steps 2-4 each time they solve a CSD problem instance.

We define the *k-shortest paths with budget constraints problem* as follows: Given a graph $G = (V, E)$, find all possible paths between start node $s$ and target node $t$ that are within the travel time/cost/budget, denoted by $B$.

It is worth noting that the budget $B$ in this section is not equivalent to the maximum willingness to detour of SPV drivers, although they are related. The maximum willingness to detour of an SPV driver is defined as the maximum time the driver is willing to delay arriving at their destination given the departure time from their origin, where delay includes waiting at the depot in addition to network routes that are not on the shortest path to the SPV driver's destination. In contrast, in this step, we generate routes for SPVs from the depot to their destinations (the blue paths in Figure 2), and therefore, the budget $B$ for an SPV $s$ equals the SPV's maximum willingness to detour ($B_s^M$) minus the shortest path travel time from the SPV's origin to the depot ($c_s^o$) and PDO pickup delay time ($\tau_{pd}$), represented as $B_s = B_s^M - c_s^o - \tau_{pd}$.

To solve the *k-shortest path with budget constraints* problem, we implement a modified version of Yen's algorithm (Yen, 1971) that includes budget constraints. Appendix A provides the pseudocode for the budgeted Yen's algorithm.

The complexity of Yen's original algorithm is $O(kV(E + V\log V)$, with $k$ being the number of paths to be generated. The budgeted Yen's algorithm keeps the heap structure of storing paths and applies Dijkstra's algorithm for shortest path finding, which is $O(E + VlogV)$. In the worst-case scenario, when the network is fully connected, and the budget is large, all the nodes along the spur path will be visited. In this case, Dijkstra's algorithm would be called $B \times |V|$ times if we treat budget $B$ as a large constant. The overall complexity for the budgeted Yen's Algorithm is $O(BV(E + V\log V)$. For a 2000 SPV case with a 10-minute depot-to-driver-destination delay budget, the computational time for the budgeted Yen's Algorithm is around 5 minutes. For a 20-minute depot-to-driver-destination delay budget in our case study, Yen's Algorithm takes roughly 48 minutes.

## 4.3 Step 2: Matching PDOs to SPV Routes

The next step in the D-H is to match the PDOs with SPVs. However, instead of matching PDOs to the SPV vehicles themselves, the D-H algorithm attempts to match PDOs to *SPV routes* that were generated from the previous step. Formulation 3 provides a formulation of the PDO-SPV route assignment problem.

### 4.3.1 Formulation 3

$$\max_{x_{prk}} \Theta_3 = \sum_{s}\sum_{r}\sum_{p} \omega a_{prs} x_{prs} - \sum_{s}\sum_{r} c_{rs} z_{rs} \tag{27}$$

subject to:

$$\sum_{s}\sum_{r} x_{prs} \leq 1 \qquad \forall p \in P \tag{28}$$

$$\sum_{r} z_{rs} \leq 1 \qquad \forall s \in S \tag{29}$$

$$\sum_{p} x_{prs} \leq z_{rs} q_s \qquad \forall (r,s) \in R \tag{30}$$

$$x_{prs}, z_{rs} \in \{0,1\} \qquad \forall p \in P, \forall (r,s) \in R \tag{31}$$

- $x_{prs} \in \{0,1\}$, Equal to one if PDO $p$ is assigned to route $r$ of SPV $s$
- $z_{rs} \in \{0,1\}$, Equal to one if SPV $s$ uses route $r$

The objective function (27) maximizes the total benefit of matching PDOs to SPV routes. To encourage successful matchings, $\omega$ (a large number) is introduced as a reward term. This operationalizes the strategy to initially match as many PDOs to SPVs as possible. In the objective function, $a_{prs}$ is a binary parameter that indicates whether the $r^{th}$ route/path of SPV $s$ can feasibly serve PDO $p$. The constraints in Eqn. (28) guarantee that a PDO is served by at most one route and at most one vehicle. The constraints in Eqn. (29) ensure each SPV only travels on at most one path through the depot. The constraints in Eqn. (30), ensure a PDO $p$ is only assigned to route $r$ if a SPV $s$ is assigned to route $r$. If $z_{rs}$ equals zero, SPV $s$ does not use route $r$, and therefore, no PDO should be served by vehicle $s$ on route $r$. Moreover, if $z_{rs}$ equals one, then vehicle $s$ does use route $r$, and the total PDOs carried by the vehicle should not exceed the maximum number of PDOs that SPV $s$ is willing to serve. Constraint (30) also acts as a linking constraint between decision variables $x_{prs}$ and $z_{rs}$.

Without Constraints (29) and the decision variable $z_{rs}$, Formulation 3 becomes a bi-partite matching problem, which is solvable in polynomial time and has a complexity of $O(n^3)$. To take advantage of the complexity of the bi-partite matching problem, we implement a Benders decomposition on Formulation 3 for large-scale cases when the SPV's willingness to detour is high. The formulation and procedure for performing Benders decomposition are as follows.

### 4.3.2 Formulation 4

**Master Problem (MP)**

$$\max_{z_{rs}} \Theta_{MP} = Z \tag{32}$$

subject to

$$\sum_{r} z_{rs} \leq 1 \qquad \forall s \in S \tag{33}$$

$$Z \leq Cuts \tag{34}$$

$$z_{rs} \in \{0,1\} \qquad \forall (r,s) \in R \tag{35}$$

**Subproblem**

$$\max_{x_{prk}} \boldsymbol{\Theta}_{\mathbf{SP}}(\bar{\boldsymbol{z}}_{\boldsymbol{rs}}) = \sum_{(r,s)} \sum_{p} \omega a_{prs} x_{prs} - \sum_{(r,s)} c_{rs} \bar{z}_{rs} \tag{36}$$

subject to

$$\sum_{(r,s)} x_{prs} \leq 1 \qquad \forall p \in P \tag{37}$$

$$\sum_{p} x_{prs} \leq \bar{z}_{rs} q_s \qquad \forall (r,s) \in R \tag{38}$$

$$x_{prs} \geq 0 \qquad \forall (r,s) \in R \tag{39}$$

The dual subproblem is formulated as follows:

$$\min_{\lambda} \boldsymbol{\Theta}_{\mathbf{DSP}}(\bar{\boldsymbol{z}}_{\boldsymbol{rs}}) = \sum_{p} \lambda_p + \sum_{(r,s)} \bar{z}_{rs} q_s \lambda_{(r,s)} - \sum_{(r,s)} c_{rs} \bar{z}_{rs} \tag{40}$$

subject to

$$\lambda_p + \lambda_{(r,s)} \geq \omega a_{prs} \qquad \forall p \in P, \forall (r,s) \in R \tag{41}$$

$$\boldsymbol{\lambda} \geq 0 \tag{42}$$

Formulation 4 presents the master problem (MP) and subproblem (SP) of the PDO-SPV route assignment problem. The subproblem is a linear assignment problem with a time complexity of $O(n^3)$. To obtain optimal cuts (i.e., corner solutions), the study solves a dual sub-problem (DSP). When initialized with a feasible solution from the master problem, the subproblem is always feasible because the linear assignment problem always has a solution given valid parameter settings. Hence, the dual subproblem is never unbounded, and one does not need to generate feasibility cuts (i.e., extreme rays) for the master problem. Therefore, the cuts in Eqn. (34) are all optimality cuts, which can be represented by line 6 in Algorithm 1. The solution procedure is a standard Benders decomposition procedure, displayed below.

| | **Algorithm 1: Benders decomposition for PDO – SPV assignment problem** |
|---|---|
| *1* | **Initialization:** |
| *2* | A feasible solution $\tilde{\mathbf{z}}^{\mathbf{0}}_{r,s}$; $LB = -\infty, UB = +\infty$. Iteration counter t = 1 |
| *3* | **While** $UB - LB > \varepsilon$: |
| *4* | **Solve** the dual subproblem (DSP) |
| *5* | Get extreme points $\tilde{\boldsymbol{\lambda}}^{\boldsymbol{t}}$ |
| *6* | Add cut $Z \leq \sum_i \bar{\lambda}^t_p + \sum_{(r,s)} \bar{z}^{t-1}_{rs} q_s \bar{\lambda}^t_{(rs)} - \sum_{(r,s)} c_{rs} \bar{z}^{t-1}_{rs}$ |
| *7* | $LB = \max\{LB, \sum_i \bar{\lambda}^t_p + \sum_{(r,s)} \bar{z}^{t-1}_{rs}[(q_s + 1)\bar{\lambda}^t_{(r,s)} - c_{rs}]\}$ |
| *8* | **Solve** the master problem |
| *9* | Get solution $UB = \Theta^*_{\mathrm{MP}}$ |
| *10* | **End While** |
| *11* | **Return**: $LB, \tilde{\mathbf{x}}_{prs}, \tilde{\mathbf{z}}_{rs}$ |

In Formulation 4, the subproblem is convex and is a restricted problem of the original problem (Formulation 3). Therefore, solving the subproblem gives a lower bound (LB) for the original problem. On the other hand, the master problem is non-convex with all integer constraints and is a relaxation of the original problem. Solving the master problem gives an upper bound (UB). Compared with the original problem, the problem size (number of constraints and variables) is significantly smaller. Adding cuts generated from the subproblem restricts the master problem gradually and reduces the gap between LB and UB.

After obtaining the initial matching results, we identify all SPVs assigned to at least one PDO, and we store their feasible routes (regardless of used or not). We call this reduced set of SPV candidate routes, $R_{reduce}$ We will use $R_{reduce}$ in Step 4 when swapping PDOs between SPVs and DVs.

## 4.4 Step 3: Assigning PDOs to DVs and Routing DVs

After solving the PDO-SPV route assignment problem, we solve the m-VRP to assign previously unassigned PDOs to DVs and to route the DVs (Step 3). To solve the problem, we employ the insertion algorithm described in Campbell and Savelsbergh (2004). The insertion heuristic has a time complexity of $O(n^3)$. Appendix B includes the pseudocode.

## 4.5 Step 4: Package Delivery Order Switching

The last step of the D-H algorithm involves an iterative procedure for switching PDOs between SPVs and DVs. Algorithm 2 in the flowchart below describes each step of the switching procedure.

We start with the initial solution generated after Steps 2 and 3, labeled as $x^{init}$. The total cost of solution $x^{init}$ is $\Theta^{init}$. We set a starting temperature of $Temp_0$, and a cooling schedule ($Step_{cool}$), which is ten percent of the current temperature, $Temp_{curr}$. The initial direction of switching is from SPV to DV. We first check the switching direction; if it is SPV to DV, we choose $n$ random PDOs to move. We keep $n$ as a small integer parameter to move to neighboring solutions; in the case study, we use $n = 1$. After switching, we obtain the candidate solution, $x^{cand}$, and the total cost of the candidate solution, $\Theta^{cand}$.

When the switching direction is from SPV to DV, we first update the assignment results of the SPVs. Since we only remove PDOs from the SPV assignments and calculate the cost reduction, we do not rematch SPVs and PDOs. We insert the PDOs moving from SPVs to DV routes using the insertion algorithm described in Step 3. We then obtain a candidate solution and can calculate the new SPV cost (small reduction) and DV cost (small increase). The total cost of the candidate solution is the summation of the DV and SPV cost components.

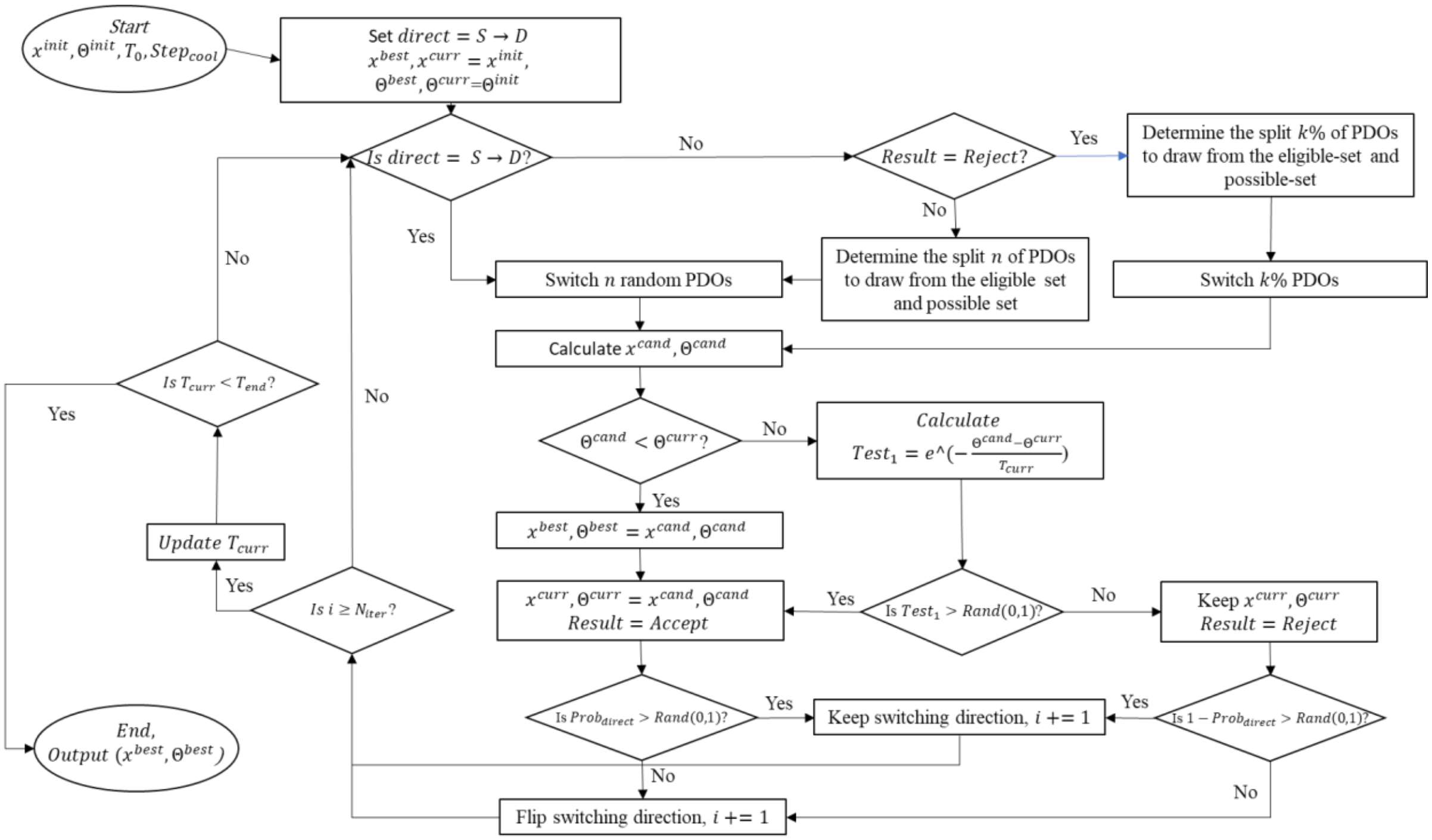

**Algorithm 2: The PDO Switching Procedure**

When the switching direction is from DV to SPV, and the number of PDOs switched is a small integer, we search through all feasible routes of unused SPVs and the existing routes of used SPVs. We then assign each PDO to the least-cost route, subject to all relevant constraints. If the number of PDOs to switch is large, e.g., more than 5 PDOs, we solve the PDO-SPV route matching problem as described in Step 2/Section 4.3. However, when solving this matching (and any PDO-SPV route matching in Step 4), we only consider the reduced SPV candidate route set, $R_{reduce}$. After the PDO-SPV assignment process, if any PDOs cannot be served by an SPV, these PDOs are inserted back into DVs using the insertion heuristic. After obtaining the SPV assignment and DV routes, we have the candidate solution and total cost.

Next, we follow the standard procedure of SA to evaluate the solution and decide whether to keep the solution. Note that we also create a result tag that is either "Accept" or "Reject" to represent the status of the current solution. After evaluating the candidate solution, we decide on the direction of switching using a probability test. We introduce another parameter $Prob_{direct}$. $Prob_{direct}$ is between 0.5 $and$ 1; in the case study, we apply 0.75. Similarly, if the candidate solution does not perform better than the current best solution, we have a $Prob_{direct}$ chance of flipping the switching direction for the next iteration.

When the direction is to move PDOs from DV to SPV, if the direction flip stems from a non-improvement in solution quality from the previous iteration, we move a large number of PDOs. The large number is a proportion ($k\%$) of all PDOs currently assigned to DVs. In the case study, we set $k\% = 10\%$. The purpose is to escape from the current state.

We do not explain the selection procedure for moving PDOs from DV to SPV in the flowchart due to space restrictions, but we will explain it herein. We first define two sets of PDOs.

**SPV-eligible PDOs**: PDOs successfully assigned to an SPV in a prior iteration.

**SPV-possible PDOs**: PDOs not assigned to an SPV in any prior iteration.

After obtaining the SPV-eligible and SPV-possible sets, we perform a Monte Carlo method to select PDOs. We first calculate the number of PDOs to switch. For every PDO to switch, we generate a random number between 0 and 1 and compare it with the $Prob_{choi}$, where:

$$Prob_{choi} = \exp\left(-\frac{Current\ Temperature}{Statrting\ Temperature}\right)$$

If $Prob_{choi} \geq rand(0,1)$, we choose a DV-served PDO that is in the SPV-eligible set; otherwise, we choose a DV-served PDO that is in the SPV-possible set. The rationale behind the selection procedure is as follows. In the early stages of annealing, $Prob_{choi}$ is smaller than 0.5 and encourages the selection of PDOs from the SPV-possible set to maximize the number of PDOs that SPVs can serve. In the late stages, $Prob_{choi}$ is closer to 1, which means that the PDOs that are in the SPV-possible set are less likely to be chosen.

In the next two sections, we conduct a real-world city-scale case study of CSD and compare the computational results from the D-H algorithm with the exact method of solving the *MFOCVRPTW*.

# 5 Numerical Case Study Overview

We conducted a numerical case study using the City of Irvine, CA, USA road network. The network contains 442 nodes and 648 links. We selected one node, displayed as a blue triangle in Figure 3, as the depot. Figure 3 displays the City of Irvine network marked with nodes, links, a depot, and PDO destinations. Each PDO has a latest permissible delivery time (time window). We randomly selected a latest delivery time value for each PDO from the following set: {12pm, 4pm, 8pm}.

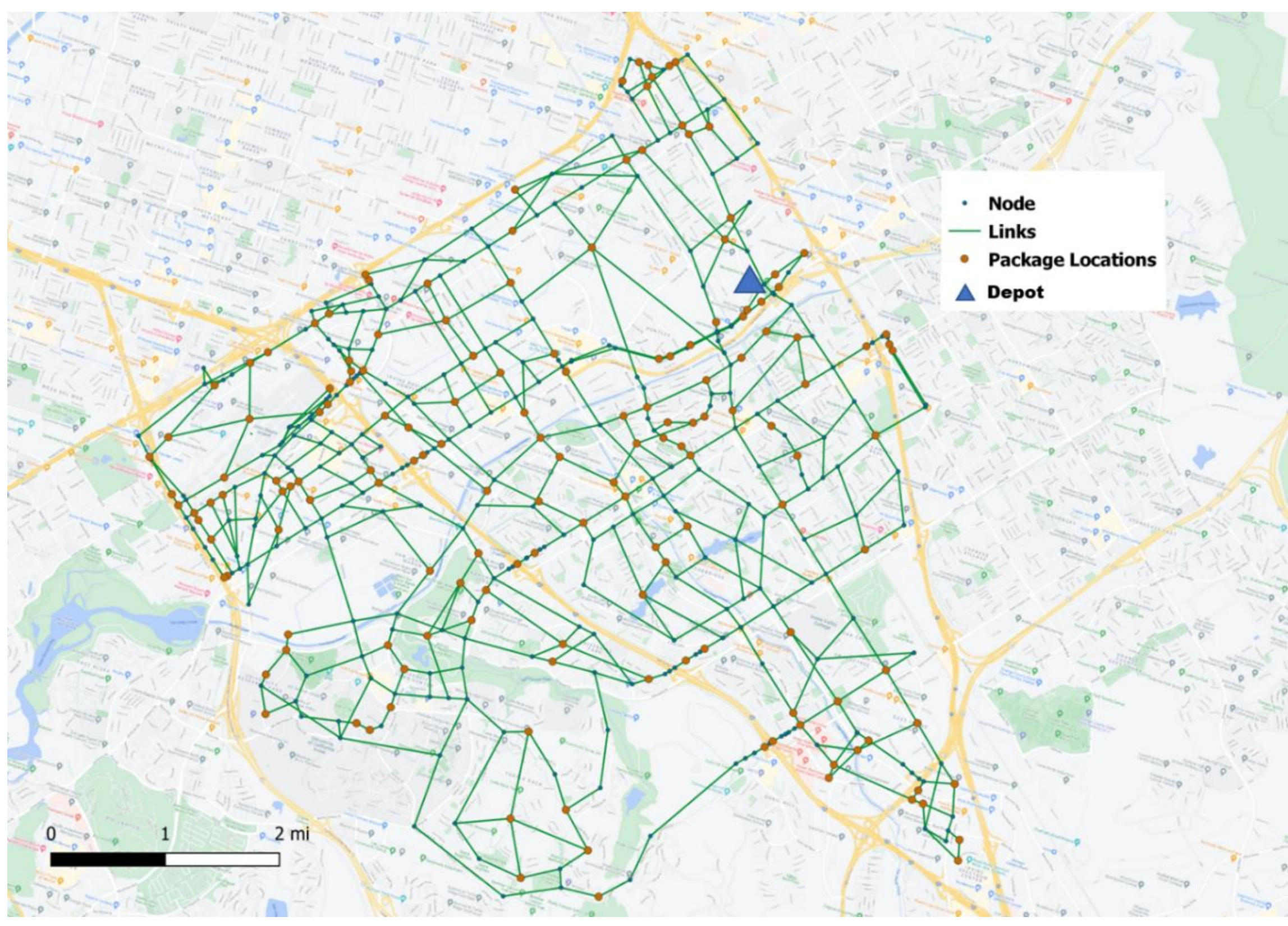

**Figure 3: Case study network structure**

Table 3 displays the relevant parameter values for the numerical case study. The case study includes 200 PDOs and a maximum of 1200 SPVs. We uniformly distributed the PDO locations across the network. Multiple packages may be in the same PDO, and multiple PDOs may share the same physical delivery location.

**Table 3: Parameter Values in the Numerical Case Study**

| Parameter | Value |
|---|---|
| Area | 32 $mile^2$ |
| Depot location | 152688 |
| Number of nodes | 442 |
| Number of PDOs | 200 |
| Number of SPVs | 0 ~ 1,200 |
| SPV average speed | 40 mph |
| SPV max stop willingness | 1 ~ 4 |
| SPV max willingness to detour | 20, 25, 30* |
| SPV detour compensation rate | $ 0.56 per mile |
| SPV PDO compensation | $ 1.5 per PDO |
| SPV PDO pickup time at depot | 10 minutes |
| DV average speed | 30 mph |
| DV per mile cost | $ 1.5 per mile |
| DV fixed cost | $ 120 per use |

* The benchmark case

We obtain SPV trips using the California State Travel Demand Model. Each SPV has a maximum number of stops that it will make. Each SPV driver has an earliest starting time (EST) from their origin and a latest arrival time (LAT) at their destination. The maximum total tour time (not the maximum tour detour time) that an SPV can afford is the difference between LAT and EST. The maximum *willingness to detour* for a given SPV $s$ is as follows: $LAT_s - EST_s - c_s$, where $c_s$ is the shortest path travel time between the trip origin and destination locations of SPV $s$.

The baseline maximum willingness to detour is 30 minutes. In the sensitivity analysis, we compare two additional willingness to detour values of 20 and 25 minutes against the base scenario of 30 minutes. We consider a waiting/pickup delay at the depot for each SPV and set the delay time at 10 minutes. This time includes loading PDOs into the SPV and entering and exiting the depot facility. When applying Yen's algorithm, we deduct this ten minutes from the total allowable depot-to-SPV destination tour time when generating SPV routes. We assume SPVs travel 40 miles per hour and DVs travel 30 miles per hour. In the base scenario, we assume that once the platform matches a PDO to an SPV driver, the driver will not reject the assignment. As a comparison, we consider possible scenarios where some drivers reject the assignment.

As mentioned, SPV drivers receive compensation for each PDO they deliver and each detour mile required to deliver those PDOs. We use the IRS's standard mileage rate for business travel, \$0.56/mile, for detour-based compensation (Internal Revenue Service, 2021). We also use \$1.50 as the compensation per PDO delivered. Based on these parameter values, the top graph in Figure 4 displays SPV driver compensation for one personal trip as a function of detour mileage and PDOs delivered.

The bottom graph in Figure 4 also displays the equivalent hourly SPV driver compensation rate, assuming an average SPV speed of 40 miles per hour and 10 minutes of delay at the depot. Ahamed et al. (2021) and Kafle et al. (2017) use an hourly rate of \$10, while Le et al. (2021), who assume CD drivers share trips with parcels, use an hourly rate of \$12. Hence, aside from SPV drivers with short detours and only one PDO to deliver, the compensation rate for SPV drivers in our study is relatively high (at least on an hourly basis).

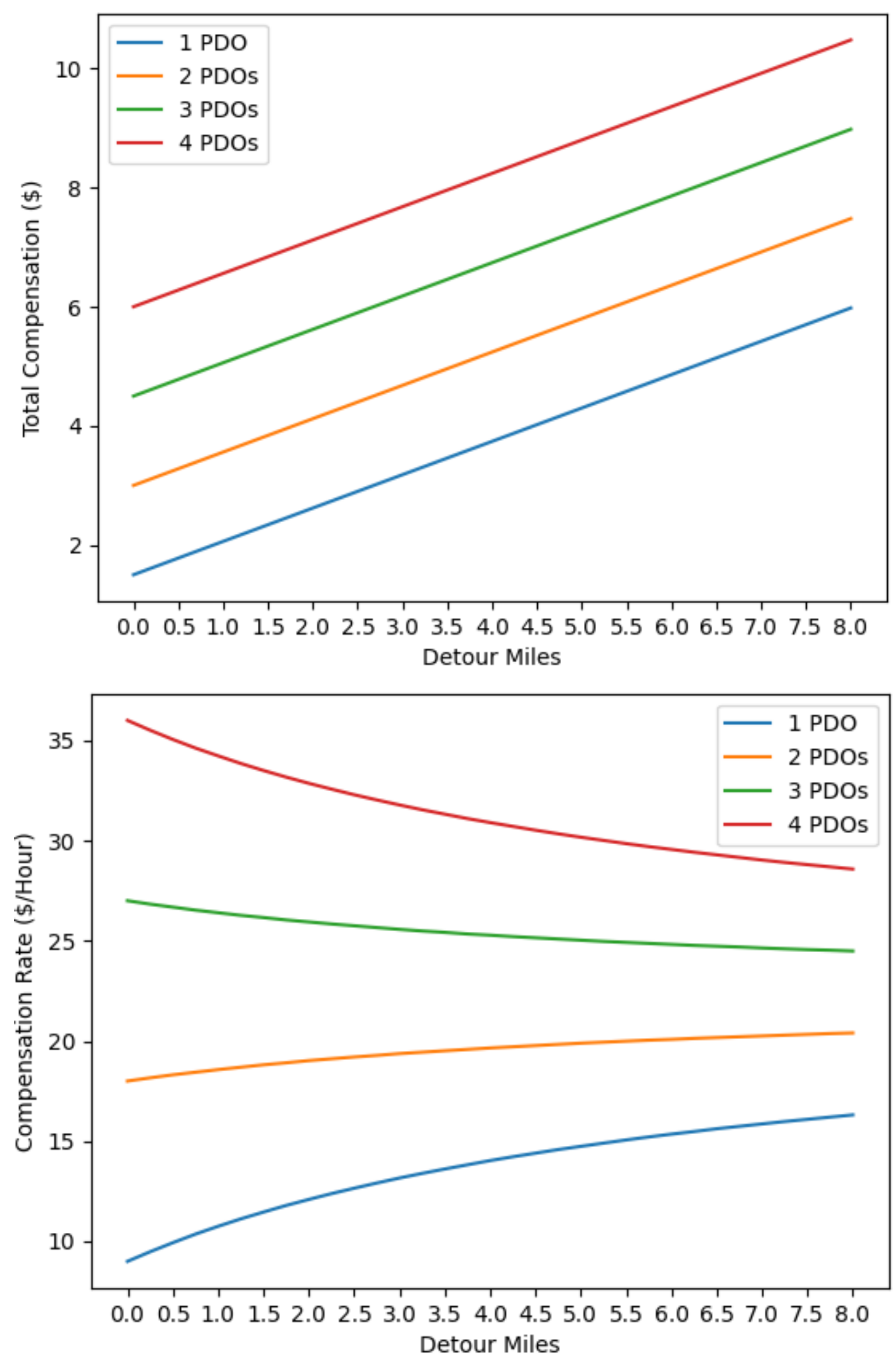

**Figure 4: Total SPV driver compensation (top) and the hourly SPV driver compensation rate (bottom) as a function of detour miles and PDOs served.**

There are also several homogenous DVs available at the depot. The DVs are responsible for delivering the PDOs not served by the SPVs. We refer to Amazon Logistics (2021) for the normal size of vans and trucks used for urban last-mile delivery, which are usually 9' to 12' vans or trucks (e.g., Ford Transit). The DV fixed cost in this study is $120, where the driver's daily salary/wage is the main cost component. The driver's hourly wage is obtained from Indeed (Indeed.com, 2023). The DV variable per-mile cost of $1.5/mile comes from the Alternative Fuel Data Center (U.S. Department of Energy, 2023), where $0.4/mile is from fuel and $1.1/mile is from depreciation, maintenance, and other minor costs.

The next section compares solving the CSD problem with the D-H and a commercial solver. The metrics used for comparison are the computational time and the optimality gap.

To evaluate the effectiveness of a CSD system, we compare the total delivery cost and vehicle miles traveled (VMT) of using CSD and the conventional DV-only delivery system. We also analyze the impact of SPV quantity by varying the number of candidate SPVs. We also conduct a sensitivity analysis on the impact of the SPV maximum willingness to detour, where we define the maximum willingness to detour as the latest time the driver is willing to arrive at their destination minus their earliest departure time.

# 6 Results and Discussion

This section compares our novel D-H algorithm to a commercial solver and an LNS algorithm regarding solution quality and computation run time (Section 6.1). Then, we use the D-H algorithm to analyze the costs and benefits of the proposed CSD system at scale under several different sets of input parameters (Section 6.2-6.7).

## 6.1 Computational experiments and comparisons

We conduct four computational experiments in this subsection. The first compares the computational results of the D-H algorithm with those of an exact solution method. The second compares the D-H algorithm with an LNS algorithm. The first two experimental comparisons contain small-scale problems (e.g., 10 PDOs, 10 SPVs) and larger-scale problems (e.g., 100 PDOs, 1000 SPVs). The third analyzes the scalability of the D-H algorithm concerning the number of PDOs. The fourth illustrates the evolution of the D-H algorithm for a large number of iterations.

We implemented the D-H in Python 3.7 and used the exact solution method in the commercial solver Gurobi (version 10.0.1). We use a 2.20 GHz Intel Xeon Server with 128 GB RAM. Since the number of PDO-switching-iterations impacts computational time and solution quality, we always use 220 as the number of iterations for the PDO-switching process.

For a complete comparison, we varied the number of PDOs from 10 to 100, where we distributed the PDOs uniformly in the study area. We also randomly generated the origins and destinations of SPVs throughout the region. In the experiments in this section, every SPV has a fixed detour time of 30 minutes. The experiments compare the computational time and the optimality gaps.

We set the computational time limit for Gurobi to 1,200 seconds. If Gurobi does not finish the optimization process after 1,200 seconds, then we report the optimality gap between the primal and dual problems, along with the optimality gap between the D-H solution and the incumbent best feasible Gurobi solutions.

### 6.1.1 D-H vs. Commercial Solver

We first compare the computational results between the commercial solver and the D-H. For small-scale problems (i.e., the first two rows of Table 4), in the four cases of 10 PDOs, the optimality gap between the D-H and the exact method is 1.2%. The D-H algorithm requires significantly less computational time. In the case of 20 PDOs, the optimality gap is less than 3% in all experiments. Hence, the D-H can obtain solutions to small problem instances that are comparable in quality to exact methods while using considerably less computational effort.

For large-scale problems (i.e., the last two rows of Table 4), the commercial solver begins to slow down considerably. For the case of 50 PDOs and 50 SPVs, the commercial solver does not obtain the optimal solution within the time limit. In fact, the D-H method obtains a better solution than the commercial solver. The computational time is also one-tenth of the time used by the exact method. For the case of 50 PDOs and 100 SPVs, the D-H also finds a better solution than the commercial solver. Moreover, for cases with more than 50 PDOs and 500 SPVs, the commercial solver cannot obtain a feasible integer solution. Conversely, the D-H algorithm can find solutions to these large problem instances in a relatively short period (i.e., around 10 minutes in nearly all cases).

#### 6.1.2 D-H Swapping Procedure vs. Large Neighborhood Search Algorithm

The second comparison is between the D-H algorithm and the LNS algorithm. When implementing the LNS, we followed the instructions in Pisinger & Ropke (2019). To make a fair comparison between D-H and LNS, we initialize the LNS with the same *initial solution* that we use for the swapping procedure in D-H. Since the LNS hyper-parameter values significantly impact the performance of heuristic algorithms, we also tune the LNS hyper-parameters, such as the stop criteria and destruction size. We keep 70% of the existing matches every iteration and destroy/reconstruct the other 30%. We apply the same stopping criteria from the swapping procedure in D-H. The number of iterations is the same for the LNS and D-H swapping procedure (220 iterations). Appendix C includes the pseudocode for the LNS algorithm.

In the 10-PDO scenarios in Table 4, we observe that both LNS and D-H produce the same solution of 137.6 (a 1.2% optimality gap) for the first two columns. This solution, 137.6, is, in fact, the initial solution to both the LNS and the D-H swapping procedure subroutines. Neither LNS nor the D-H swapping procedure improves the initial solution, which indicates the quality of the initial solution obtained from Steps 1 – 3 in the D-H algorithm. In the cases with 50 and 100 SPVs (i.e., the last two columns of the first row in Table 4), the LNS improves the solution, while the D-H swapping procedure keeps the initial solution. However, LNS takes a relatively longer time than the D-H swapping procedure. In fact, in all the computational experiments, LNS results in a longer CPU time than D-H. In some scenarios, LNS almost doubles the computational time of the D-H algorithm (e.g., the case with 50 PDO and 500 SPVs). The reason is that the LNS "destroy and repair procedure" ruins 30% of all matches and requires rematching every iteration. On the other hand, the D-H swapping procedure only executes the destroy and repair procedure after several iterations of "non-improvement behavior," and the percentage to destroy and repair is only 10%.

In large-scale scenarios (i.e., the last two rows in Table 4), the D-H swapping procedure outperforms the LNS algorithm in 5 out of 8 experiments regarding solution quality. While LNS generates better solutions for the other 3, the difference is only 0.8% to 1.2%. Again, the computational time required for LNS is much longer than D-H for the large-scale instances.

We conclude that the D-H swapping procedure outperforms LNS when considering solution quality and run time jointly. We also want to note that the entire LNS depends on a quality initial solution obtained from Steps 1-3 of the D-H algorithm. If LNS were to start from a random feasible solution, it would likely take even longer to find solutions of a similar quality.

Lastly, we note that metaheuristics such as LNS depend on parameter tuning. More parameter tuning could improve both the LNS and each of the steps of the D-H algorithm.

**Table 4: Computational Experiments Comparing D-H, LNS, and Commercial Solver (Solver)**

| No of PDOs | | Solver | LNS | D-H | Solver | LNS | D-H | Solver | LNS | D-H | Solver | LNS | D-H |
|---|---|---|---|---|---|---|---|---|---|---|---|---|---|
| | ***SPVs*** | | ***10*** | | | ***20*** | | | ***50*** | | | ***100*** | |
| **10** | *Time (s)* | 1.6 | 1.4 | 1.1 | 2.3 | 1.8 | 1.1 | 10.5 | 2.5 | 1.2 | 65.9 | 4.2 | 1.3 |
| | *Cost* | 136 | 137.6 | 137.6 | 136 | 137.6 | 137.6 | 136 | 136.6 | 137.6 | 136 | 137.2 | 137.6 |
| | *OPT Gap* | | 1.20% | 1.20% | | 1.20% | 1.20% | | 0.44% | 1.20% | | 0.88% | 1.20% |
| | ***SPVs*** | | ***20*** | | | ***40*** | | | ***100*** | | | ***200*** | |
| **20** | *Time (s)* | 130.6 | 10.7 | 6.54 | 308.9 | 12.3 | 6.8 | 455.2 | 22.8 | 16.9 | 1105.6 | 31.6 | 25.3 |
| | *Cost* | 142.4 | 144.5 | 144.5 | 142.4 | 143.8 | 144.5 | 139.2 | 141.4 | 139.7 | 138.2 | 140.2 | 139.7 |
| | *OPT Gap* | | 2.70% | 2.70% | | 0.98% | 2.70% | | 1.58% | 0.40% | | 1.45% | 1.10% |
| | ***SPVs*** | | ***50*** | | | ***100*** | | | ***250*** | | | ***500*** | |
| | *Time (s)* | 1200 | 161 | 126 | 1200 | 210 | 153 | 1200 | 287 | 193 | 1200 | 403 | 226 |
| **50** | *Cost* | 194.5 | 189.3 | 191.6 | 244.5 | 190.4 | 191.2 | --- | 187.2 | 185.1 | --- | 182.6 | 181.9 |
| | *OPT Gap* | | | -1.5% (13.6%) * | | | -19.5% (35.4%) * | | | | | | |
| | ***SPVs*** | | ***100*** | | | ***200*** | | | ***500*** | | | ***1000*** | |
| **100** | *Time (s)* | 1200 | 722 | 493 | 1200 | 855 | 527 | 1200 | 983 | 586 | 1200 | 1127 | 613 |
| | *Cost* | --- | 335.1 | 327.4 | --- | 322.8 | 322.3 | --- | 282.7 | 283.2 | --- | 251.1 | 246.4 |
| | *OPT Gap* | | | | | | | | | | | | |

1. The gap between the D-H/LNS method and an exact method if Gurobi outputs a solution in 1200 secs.

2. The optimality gap and the dual gap (in parenthesis) between the primal and dual problem solutions, if Gurobi does not finish in 1200 secs. (Marked with *)

3. Not applicable if Gurobi does not start cutting planes within 1200 secs (Marked with --- ).

### 6.1.3 D-H Scalability with respect to Delivery Tasks

To test how the computational time changes as the number of PDOs increases, we fix the number of SPVs at 800 and vary the number of PDOs from 100 to 800. We also set the number of D-H iterations to 200. Figure 5 displays the computational results.

Figure 5 shows that the algorithm scales well when the number of PDOs is smaller than 600. However, when the number of PDOs exceeds 600, the computational time increases rapidly. After 800 PDOs, the run time exceeds the preset computational time limit of 3 hours.

Combining Table 4 and Figure 5, we conclude that the D-H scales better with the number of SPVs than the number of PDOs. However, we want to note that as the density of PDOs increases, it is possible to start grouping PDOs together before determining vehicle-PDO matches and vehicle routes. The impact of this grouping approach on solution quality will decrease as PDO density increases within a region.

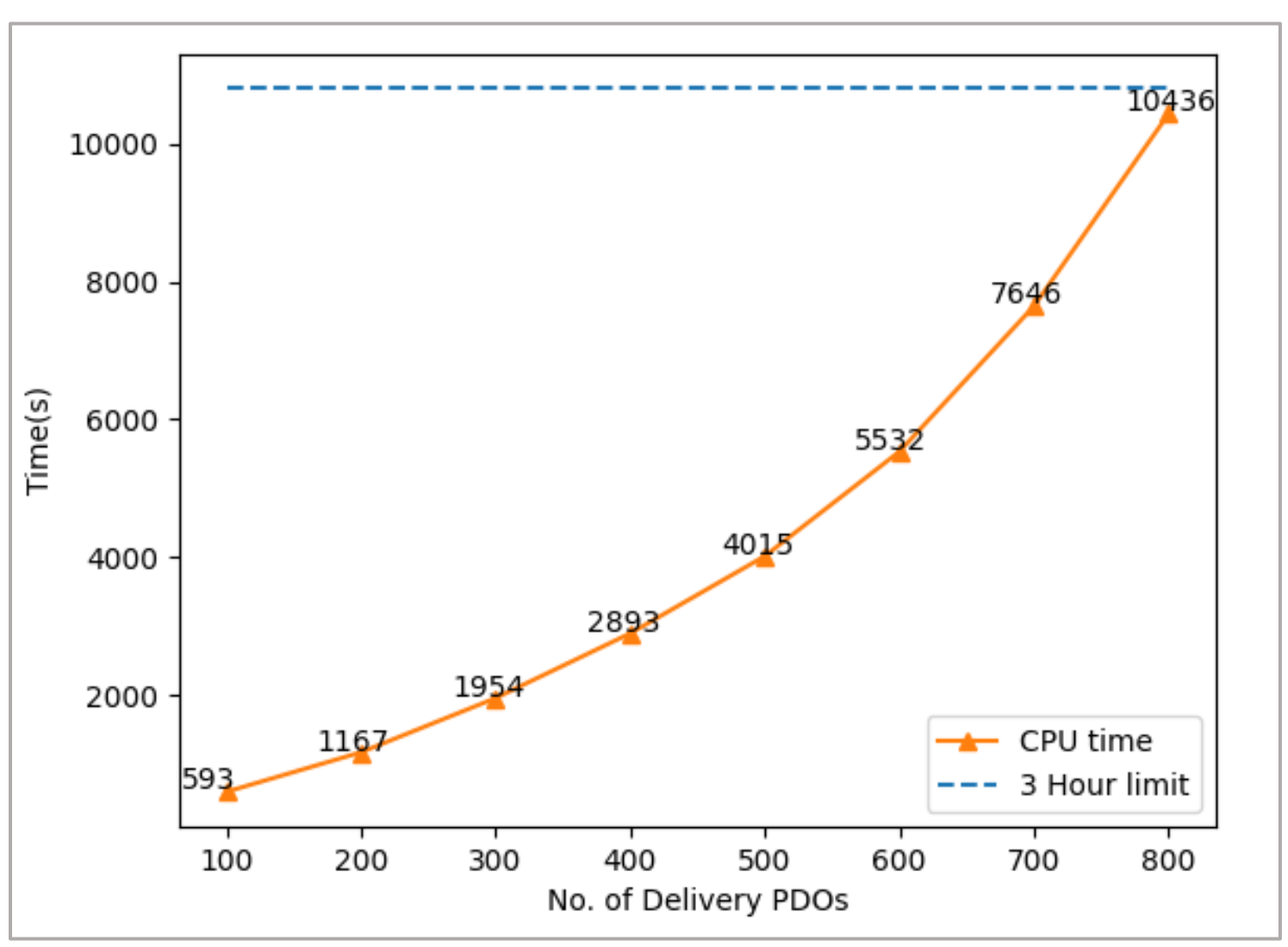

**Figure 5: Computational time as a function of the number of PDOs**

### 6.1.4 Insights into D-H Search

To illustrate how the D-H searches for solutions, we visualize a large-scale problem instance with 200 PDOs and 1200 SPVs. We run the D-H algorithm for 350 iterations, which lasts around 2 hours. Figure 6 shows how the solution changes from one iteration to the next.

Figure 6 shows that the algorithm finds a relatively good solution (objective value of 477.5) in the first 50 iterations, which takes about 20 mins. Then, the algorithm gradually improves the solution until about iteration 80, where the objective value is 469.2. After that, the solution quality fluctuates significantly until iteration 160 (466.9), wherein the algorithm finds the best solution over the first two hours.

There are two reasons for the large spikes in Figure 6. One reason is that when additional PDOs move from SPV to DV, an additional DV is required to maintain problem feasibility related to vehicle capacity or route duration. As the solution adds a DV, the cost drastically increases. Another reason is the batch movement of PDOs from DVs to SPVs. In both cases, after the large spike, the algorithm gradually improves the solution quality.

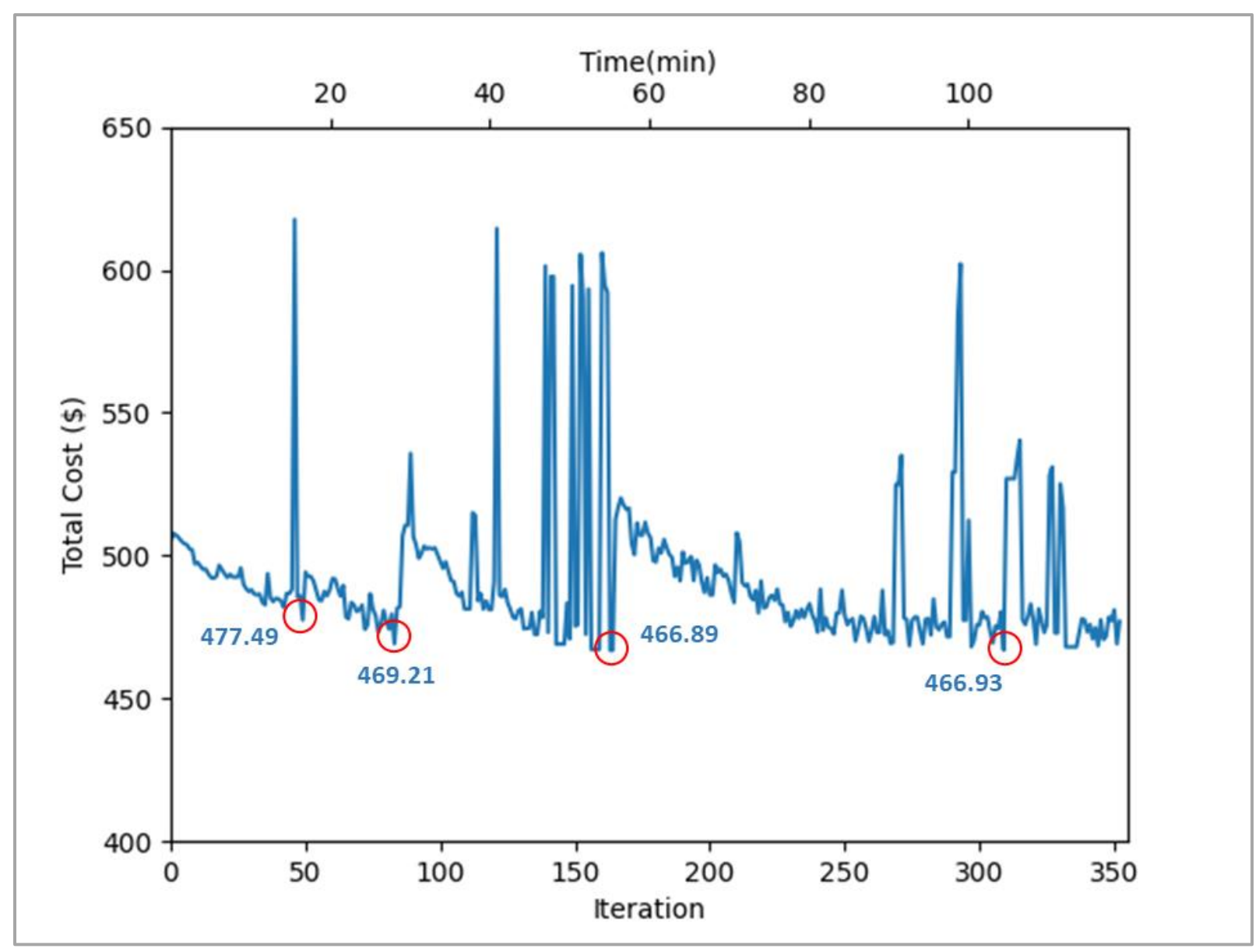

**Figure 6: Trend in solutions**

#### 6.1.5 Concluding Statement on Computational Efficiency and Solution Quality

In summary, the D-H algorithm effectively identifies high-quality solutions while solving large-scale problems quickly. For small-scale problems, the algorithm achieves near-optimal solutions in a fraction of the time of an exact solution method. For medium-scale problems, the D-H outperforms a commercial solver within the specified time limit. For large-scale problems, when the commercial solver cannot even obtain a feasible solution, the D-H obtains solutions very quickly.

The comparison between D-H and LNS indicates that D-H and LNS are comparable in terms of solution quality. However, D-H is superior in terms of computational run time.

### 6.2 Package Delivery Orders Served by SPVs

This subsection analyzes the number of PDOs served by SPVs as a function of the number of SPVs available. According to Section 4, the PDOs are matched to the SPVs first (an initial matching), and then PDOs are switched from SPVs to DVs (final matching). Figure 7 displays the initial and final matching numbers against the number of SPVs available.

The blue line in Figure 7 represents the initial number of PDOs matched with SPVs. Given the algorithm's structure, the initial matching number is the largest possible number of PDOs SPVs can serve. Unsurprisingly, the blue dashed line shows that the number of PDOs that SPVs can serve increases with the number of SPVs. Similarly, the orange line shows that the final number of PDOs served by SPVs increases with the number of SPVs in the fleet. Moreover, the dashed blue line is an upper bound on the orange line.

Even when the number of available SPVs reaches a relatively high level, SPVs do not serve all PDOs (i.e., the matching rate is still around 70%). If the logistics platform can keep increasing the number of SPVs, they can likely serve all the PDOs with SPVs. However, the total number of SPVs required may be exceedingly high. Hence, in nearly all cases, the problem instance requires at least one DV to serve PDOs.

The orange line in Figure 7 represents the number of SPV-served PDOs at the optimal or near-optimal cost. There is a stepwise-like increase in the trend. The green line indicates an increasing stepwise trend and represents the number of PDOs served by DVs. These findings indicate that the optimal solution involves utilizing DVs to full capacity (with a full load or a maximum travel time) while minimizing the number of DVs in operation. When the number of SPVs increases to a level where the feasible SPVs can serve an additional truckload of PDOs, the number of DVs required to PDOs decreases by one, and the total cost decreases.

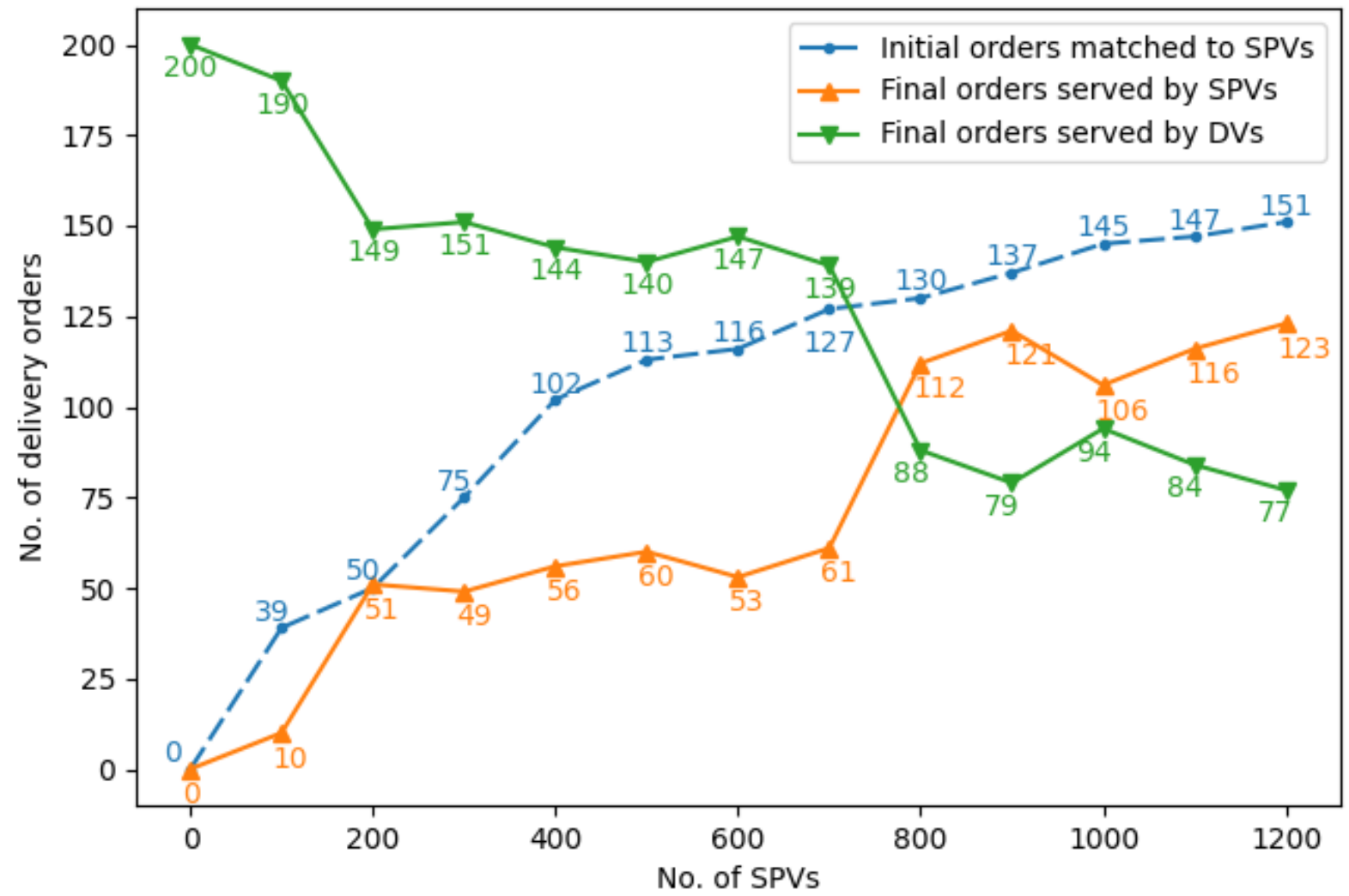

**Figure 7: PDOs served by SPVs and DVs**

### 6.3 SPV Usage Analysis

This section analyzes the SPV (route) usage from the supply perspective. Consider an SPV that travels directly from its origin to the depot on the shortest path and then travels from the depot to its destination on the shortest depot-to-destination path while delivering at least one PDO. Since the SPV does not detour from the depot to its destination, we call this a minimum depot-to-destination detour route.

Figure 8 displays a histogram of the depot-to-destination detour distances for SPVs assigned to serve PDOs. The histogram values come from computational experiments in Figure 7, where the number of SPVs ranges from 100 to 1200. In Figure 8, on the x-axis, “0” indicates a minimum depot-to-destination detour route, and the other integer numbers are the travel time differences between the used route and the minimum depot-to-destination detour route.

The average total depot-to-destination detour distance for SPVs with delivery tasks is 0.3 miles. The left-most bar in the histogram shows that 30% of SPVs use the minimum depot-to-destination detour route. Moreover, 65% of the SPVs traverse a route from the depot with no more than 2 minutes of detour. From the PDO perspective, SPVs deliver 57% of PDOs on a minimum depot-to-destination detour route. Almost 90% of the SPVs use a route with a depot-to-destination detour of less than 7 minutes.

Unsurprisingly, the histogram of used SPV routes looks similar to a negative exponential distribution because the objective function favors SPV routes with shorter detours. However, the distribution is non-monotonic concerning depot-to-destination detour time. After decreasing steadily from 10 to 14 minutes, the frequency of orders steadily increases from 15 to 18 minutes of depot-to-destination detour time. One possible reason for this phenomenon is that longer detour routes serve multiple PDOs or PDOs in the region's periphery that do not fit neatly into DV routes. As such, the phenomenon may be due to the structure of the particular service region.

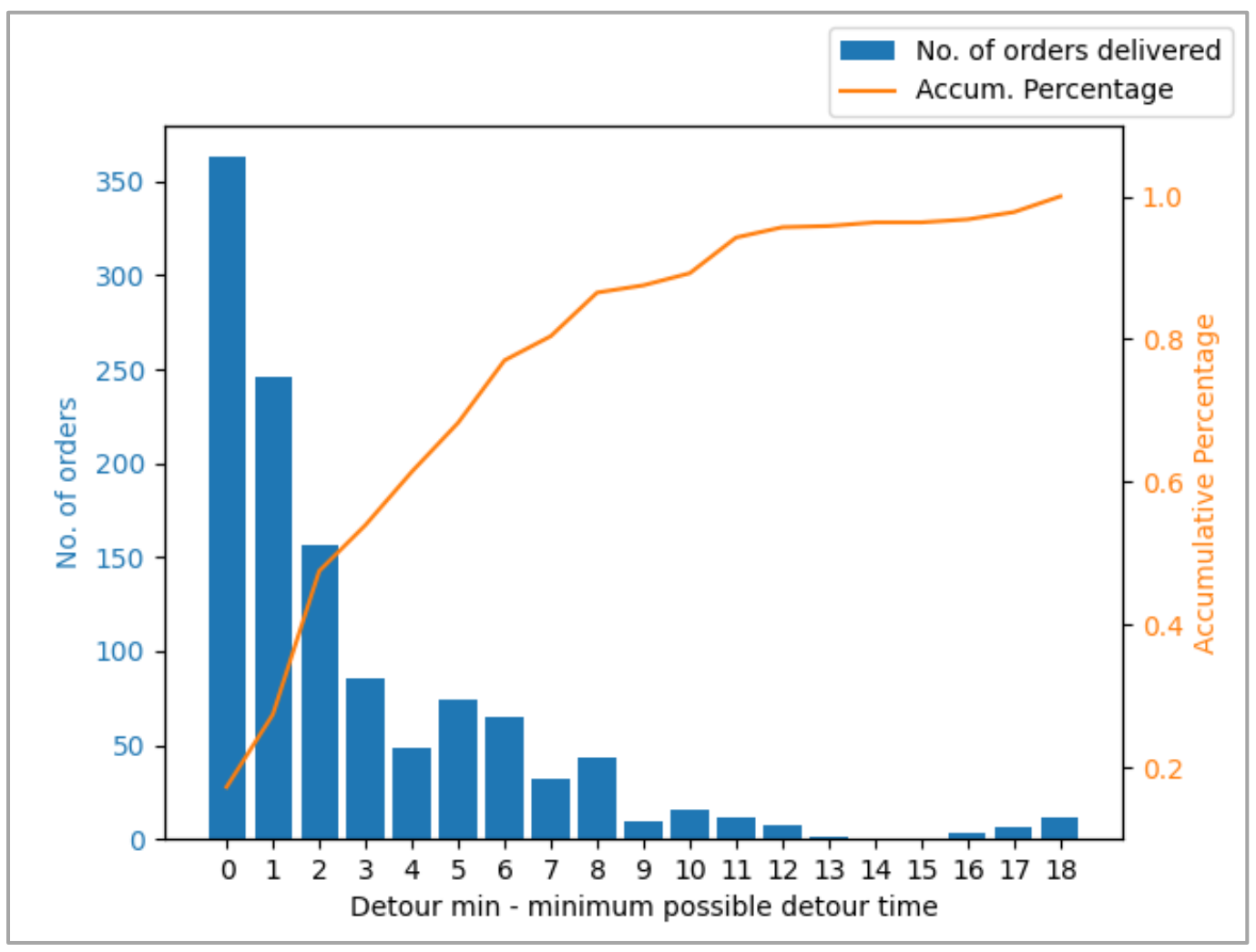

**Figure 8: Histogram of depot-to-destination detour distance for PDOs served by SPV routes**

The implications of the results in Figure 8 are two-fold. First, from a service design perspective, increasing the maximum willingness to detour is unlikely to increase the number of PDOs served by SPVs. Second, algorithmically, the budget in the budgeted k-shortest path algorithm can be small and not significantly impact the solution quality. Reducing the k-shortest path budget can significantly reduce computational complexity. Of course, the results in Figure 8 depend on the network structure and spatial distribution of PDOs and SPV trips.

## 6.4 Total Delivery Cost

Total delivery cost is the most critical metric as it is the objective function for the original problem formulation. Figure 9 displays SPV, truck, and total delivery costs as a function of the number of SPVs available. Figure 10 displays the per-PDO costs for SPVs, DVs, and all vehicles.

Figure 9 and Figure 10 illustrate that the total delivery cost (and, by definition, average cost) decreases as the number of available SPVs increases. This finding is unsurprising given that more SPVs represent a larger feasible region or more options for SPVs to serve PDOs. Consistent with the results in Figure 7, the total delivery cost in Figure 9 reduces in steps as the number of SPVs increases. This result, again, stems from reducing the number of DVs necessary to deliver the PDOs and the quality of the SPV routes.

The blue SPV cost line in Figure 9 increases with more SPVs, while the DV (orange line) and total cost (green line) decrease with the number of available SPVs. SPV costs increase with the number of SPVs due to the increase in PDOs served by SPV, while DV costs decrease due to the decrease in the number of DVs needed to serve PDOs. The results in Figure 9 indicate that the primary savings from CSD mainly stem from reducing the fixed costs of DVs.

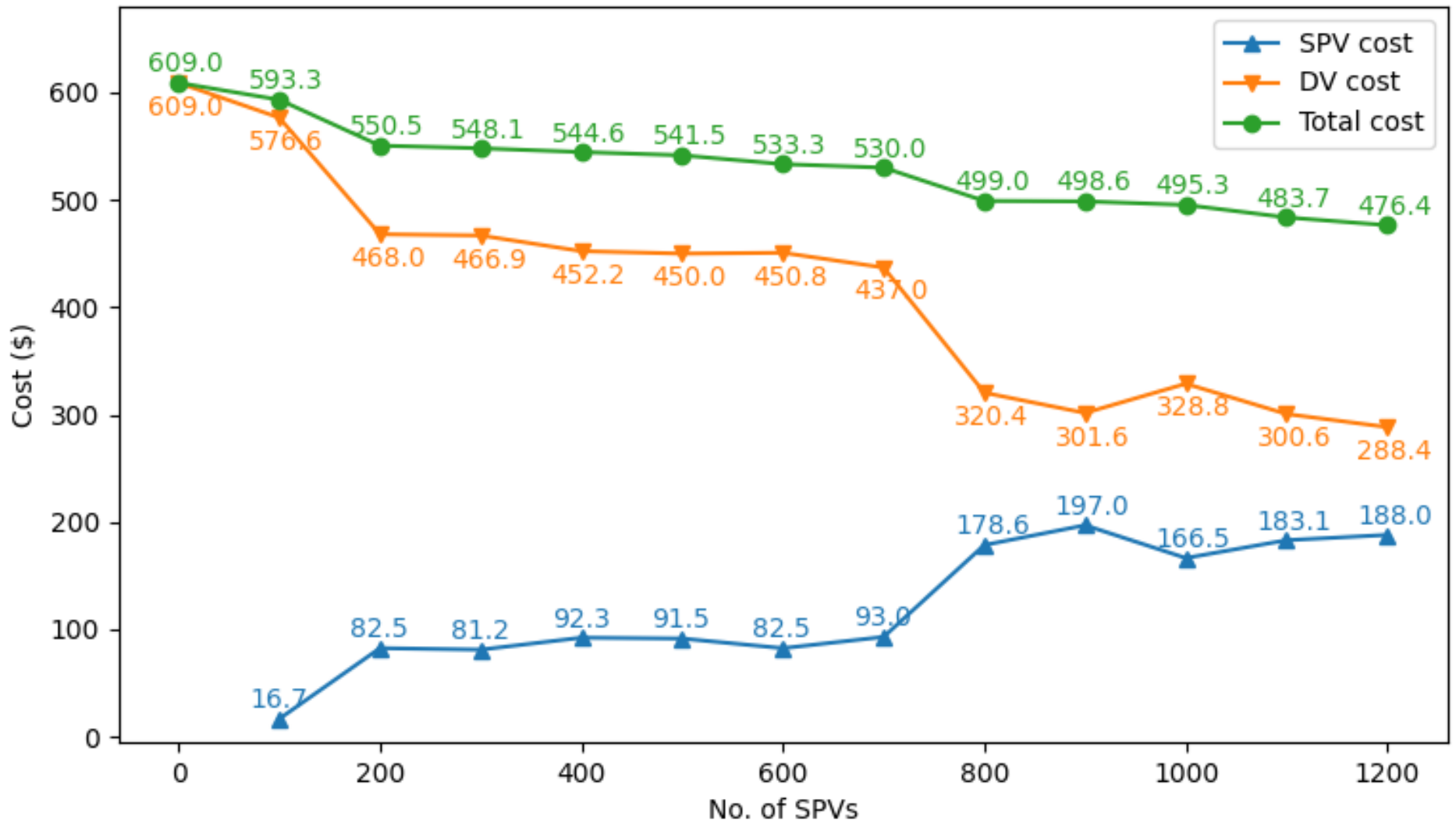

**Figure 9: Total cost and cost by vehicle type**

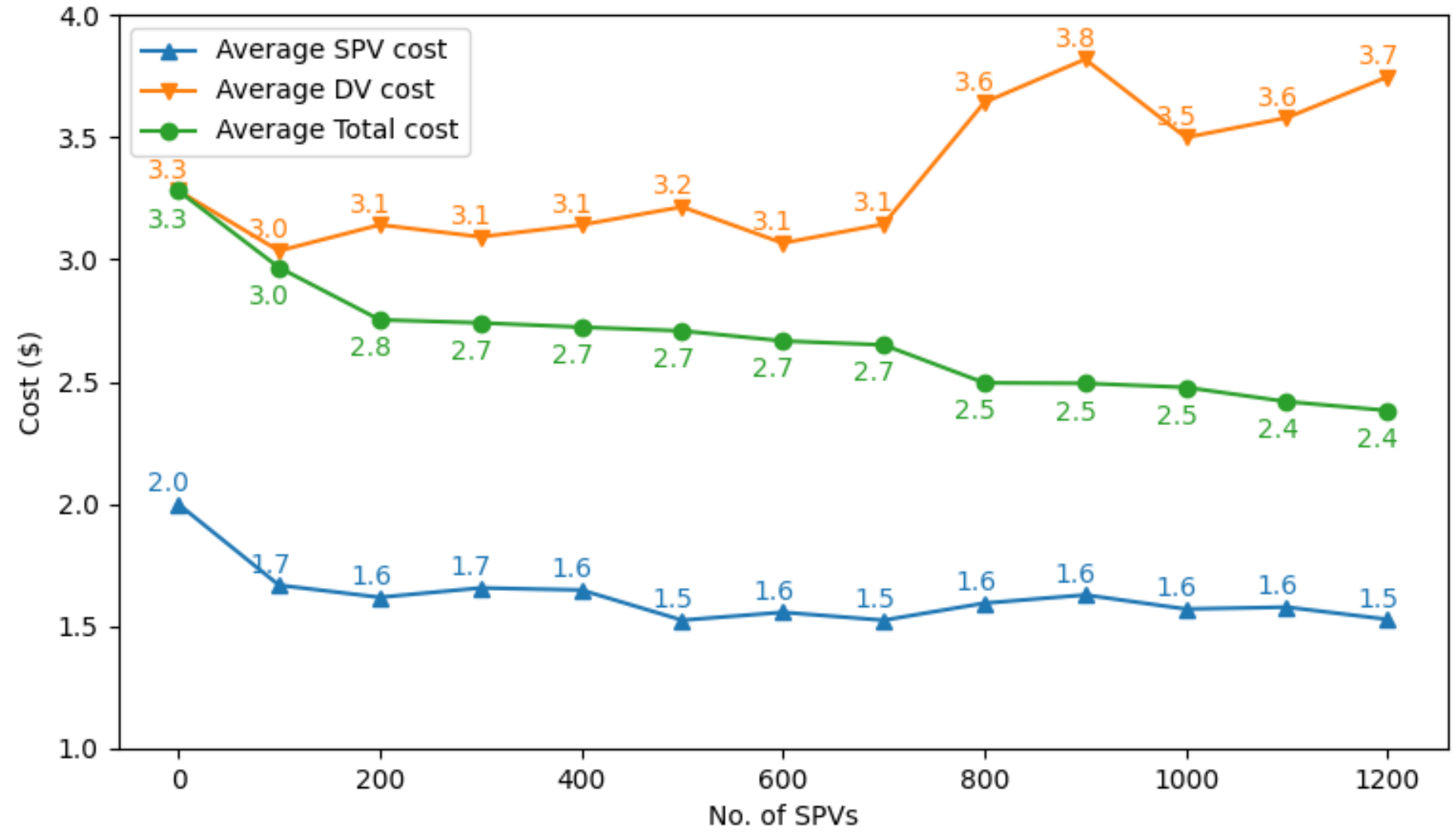

**Figure 10: Average cost per PDO**

Table 5 shows that the total cost slowly decreases when available SPVs increase from 400 to 700. The decrease stems from more SPVs allowing better matching (i.e., shorter detour distances) between SPVs and PDOs. However, compared to the decreases between 200-300 and 700-800 when the DV fleet size decreases by one in both cases, the decrease in cost between 400-700 is relatively small.

Table 5 shows that CSD significantly reduces cost compared to an all-DV fleet. The cost savings range from 10% to 22% as the number of SPVs available increases.

Besides the total cost, the average cost of delivering a PDO decreases as the number of available SPVs increases. Both Figure 10 and Table 5 demonstrate the trend. It is worth noting that while the overall average delivery cost decreases with available SPVs, the average DV delivery cost increases slightly, and the average SPV delivery cost decreases. However, because the average SPV delivery cost is always lower than the average DV delivery cost, moving more PDOs with SPVs decreases the overall average delivery cost.

**Table 5: Total and Average Delivery Cost under Different Numbers of Available SPVs**

| | PDOs served by | | Cost (S) | | | | AVG cost per PDO ($) | | |
|---|---|---|---|---|---|---|---|---|---|
| SPV Number | SPV | DV | SPV | DV | Total | % Saving w.r.t VRP | by SV | by DV | overall |
| 0 | 0 | 200 | 0 | 609 | 609.00 | - | - | 3.05 | 3.05 |
| 100 | 10 | 190 | 16.68 | 576.6 | 593.28 | 2.58% | 1.67 | 3.03 | 2.97 |
| 200 | 51 | 149 | 82.55 | 468 | 550.55 | 9.60% | 1.62 | 3.14 | 2.75 |
| 300 | 49 | 151 | 81.185 | 466.88 | 548.07 | 10.01% | 1.66 | 3.09 | 2.74 |
| 400 | 56 | 144 | 92.313 | 452.25 | 544.56 | 10.58% | 1.65 | 3.14 | 2.72 |
| 500 | 60 | 140 | 91.511 | 450 | 541.51 | 11.08% | 1.53 | 3.21 | 2.71 |
| 600 | 53 | 147 | 82.532 | 450.75 | 533.28 | 12.43% | 1.56 | 3.07 | 2.67 |
| 700 | 61 | 139 | 93.028 | 437 | 530.03 | 12.97% | 1.53 | 3.14 | 2.65 |
| 800 | 112 | 88 | 178.607 | 320.43 | 499.04 | 18.06% | 1.59 | 3.64 | 2.50 |
| 900 | 121 | 79 | 197.028 | 301.6 | 498.63 | 18.12% | 1.63 | 3.82 | 2.49 |
| 1000 | 106 | 94 | 166.504 | 328.81 | 495.31 | 18.67% | 1.57 | 3.50 | 2.48 |
| 1100 | 116 | 84 | 183.092 | 300.6 | 483.69 | 20.58% | 1.58 | 3.58 | 2.42 |
| 1200 | 123 | 77 | 188.019 | 288.4 | 476.42 | 21.77% | 1.53 | 3.75 | 2.38 |

## 6.5 Total Delivery Vehicle Mileage

This section analyzes total vehicle miles traveled (VMT), a proxy for congestion, energy consumption, and environmental impact. Figure 11 and Table 6 present the total VMT resulting from delivery activities for SPVs, DVs, and all vehicles.

The results indicate that total VMT tends to increase with the number of SPVs, although non-monotonically. This finding is unsurprising given that DV routes are more VMT-efficient than most SPV routes, and as the number of available SPVs increases, the number of PDOs assigned to SPVs increases.

Given that VMT increases with the number of SPVs, the question is whether the CSD system increases congestion and worsens the environmental impacts of PDO delivery compared to DC-only delivery. In the case of congestion, the answer is almost certainly "yes." To analyze the impact of the CSD system from an environmental perspective, we used $CO_2$ emissions to compare DV-only and a CSD system (with different SPV numbers). Figure 12 in Section 6.6 displays the results.

Also worth mentioning is that the VMT estimate for each SPV in Figure 11 includes SPV travel from the SPV driver's origin to the depot. If the logistics company only considers SPVs from people already shopping at the store (at or near the depot), which is the case in most previous studies (Archetti et al., 2016; Arslan et al., 2019; Dayarian and Savelsbergh, 2020), then total VMT is likely to decrease with the number of SPVs.

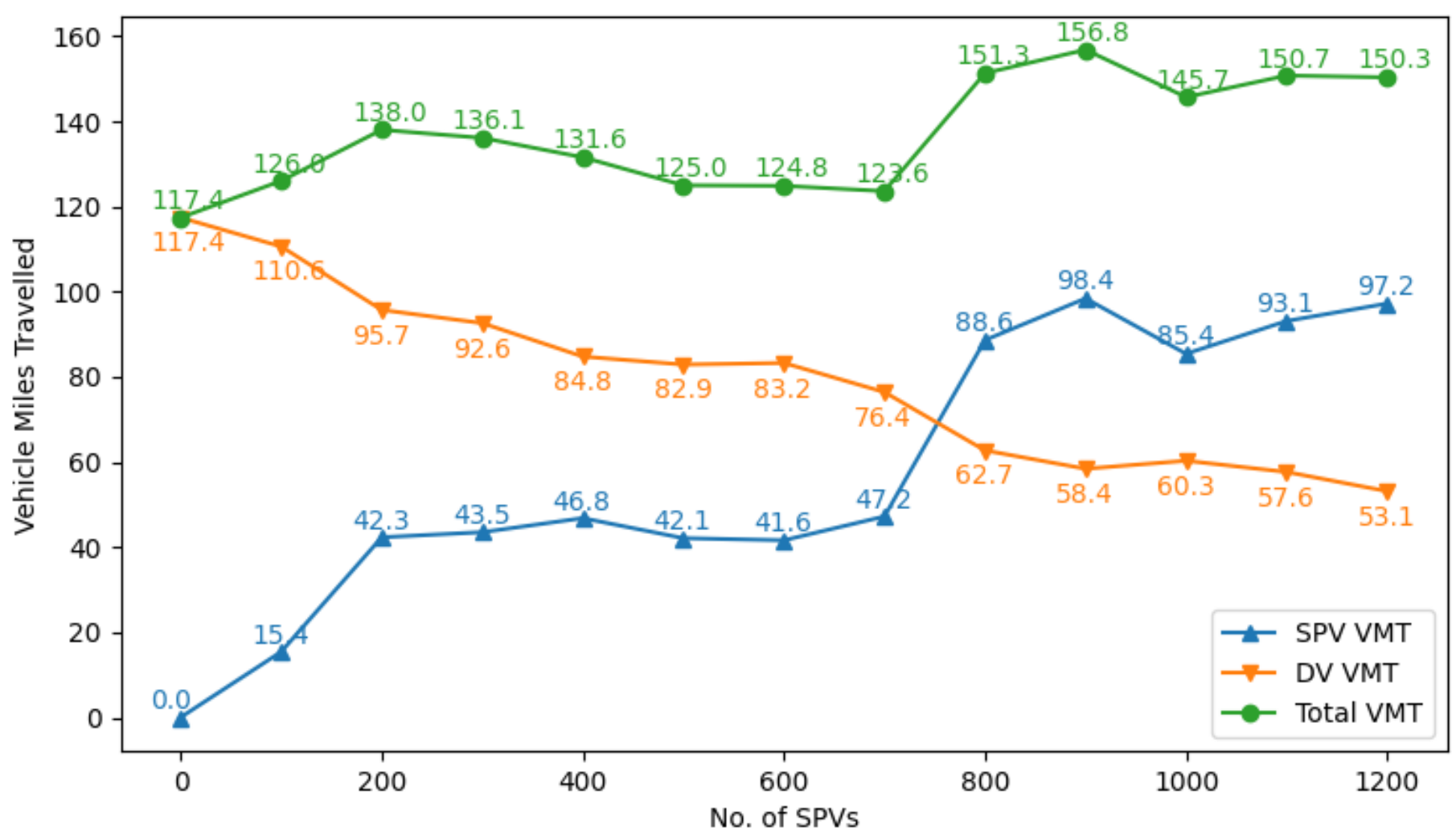

**Figure 11: VMT from PDO delivery**

**Table 6: Total and Average VMT as a Function of Available SPVs**

| | **PDOs served by** | | **VMT (miles)** Varying Numbers | | | | **AVG VMT per PDO (miles)** | | |
|---|---|---|---|---|---|---|---|---|---|
| SPV Number | SPV | DV | SPV | DV | Total | % Saving w.r.t VRP | by SV | by DV | overall |
| 0 | 0 | 200 | 0 | 117.375 | 117.38 | | - | 0.59 | 0.59 |
| 100 | 10 | 190 | 15.4 | 110.6 | 126.00 | -7.35% | 1.54 | 0.58 | 0.63 |
| 200 | 51 | 149 | 42.3 | 95.7 | 138.00 | -17.57% | 0.83 | 0.64 | 0.69 |
| 300 | 49 | 151 | 43.5 | 92.6 | 136.10 | -15.95% | 0.89 | 0.61 | 0.68 |
| 400 | 56 | 144 | 46.8 | 84.75 | 131.55 | -12.08% | 0.84 | 0.59 | 0.66 |
| 500 | 60 | 140 | 42.1 | 82.875 | 124.98 | -6.47% | 0.70 | 0.59 | 0.62 |
| 600 | 53 | 147 | 41.6 | 83.24 | 124.84 | -6.36% | 0.78 | 0.57 | 0.62 |
| 700 | 61 | 139 | 47.2 | 76.4 | 123.60 | -5.30% | 0.77 | 0.55 | 0.62 |
| 800 | 112 | 88 | 88.6 | 62.7 | 151.30 | -28.90% | 0.79 | 0.71 | 0.76 |
| 900 | 121 | 79 | 98.4 | 58.4 | 156.80 | -33.59% | 0.81 | 0.74 | 0.78 |
| 1000 | 106 | 94 | 85.4 | 60.28 | 145.68 | -24.12% | 0.81 | 0.64 | 0.73 |
| 1100 | 116 | 84 | 93.1 | 57.64 | 150.74 | -28.43% | 0.80 | 0.69 | 0.75 |
| 1200 | 123 | 77 | 97.2 | 53.125 | 150.33 | -28.07% | 0.79 | 0.69 | 0.75 |

## 6.6 Impacts of Shared Vehicle Willingness-to-Detour

This section aims to assess the impact of SPV willingness to detour. In prior sections, the baseline value for maximum willingness to detour is 30 minutes for SPVs. This section considers cases where the maximum willingness-to-detour parameter values are 20 and 25 minutes.

Figure 12 shows the impact of the maximum willingness-to-detour time on the total delivery cost. The results are consistent with expectations; the total delivery cost decreases as the maximum willingness to detour increases. The impact is significant under certain conditions regarding the number of available SPVs. At 500 available SPVs, the total delivery cost is $557 for the 20-minute maximum detour case, whereas the

total delivery costs are around $539 for the other two cases. Looking at a value of $500 on the y-axis in the top figure, the 30-minute case requires only 300 available SPVs to achieve a total cost of around $550, while the 20-minute case needs at least 700 available SPVs. On the other hand, the differences between 25-minute and 30-minute cases are insignificant. Combining the finding from Section 6.3 that most SPV drivers only incur short detours with this finding, we can conclude that from a service perspective, 25 minutes is a sufficient willingness-to-detour value for the logistics provider in this service region.

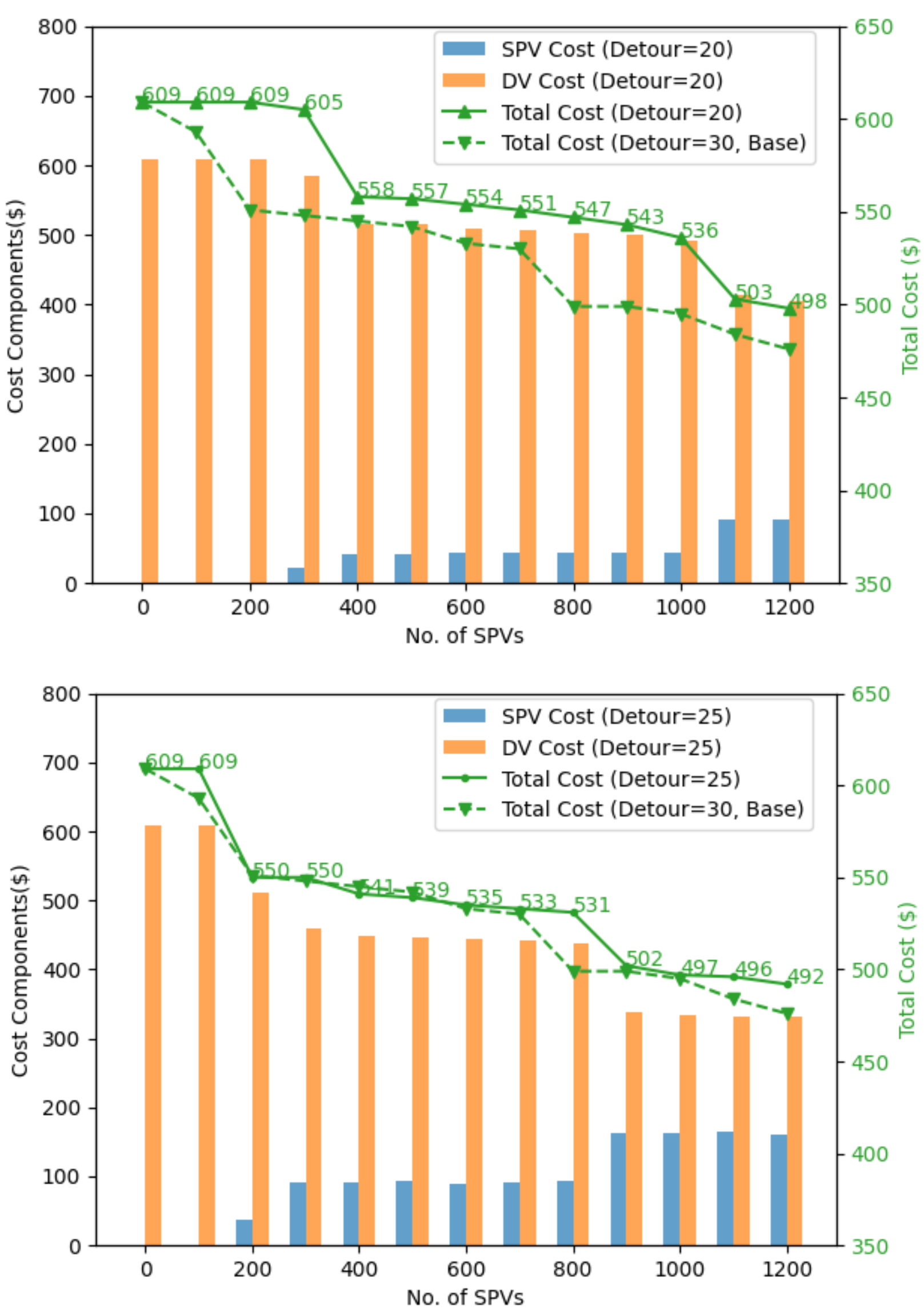

**Figure 12: Impact of Willingness to Detour on Costs. The top figure compares 30- vs. 20-min maximum detour. The bottom figure compares 30- vs 25-min maximum detour.**

Figure 13 displays the relationship between the number of PDOs that SPVs can serve (y-axis) and the total number of available SPVs (x-axis) under three different maximum willingness to detour values (i.e., the different colored lines). Figure 13 shows that with the same number of SPVs, the 30-minute detour can serve 4 to 6 times more PDOs than the 20-minute case. The results indicate that, at least for the Irvine network and the spatial demand in the scenarios in this study, there is a strong business case for incentivizing drivers to accept a 25-minute detour rather than a 20-minute detour.

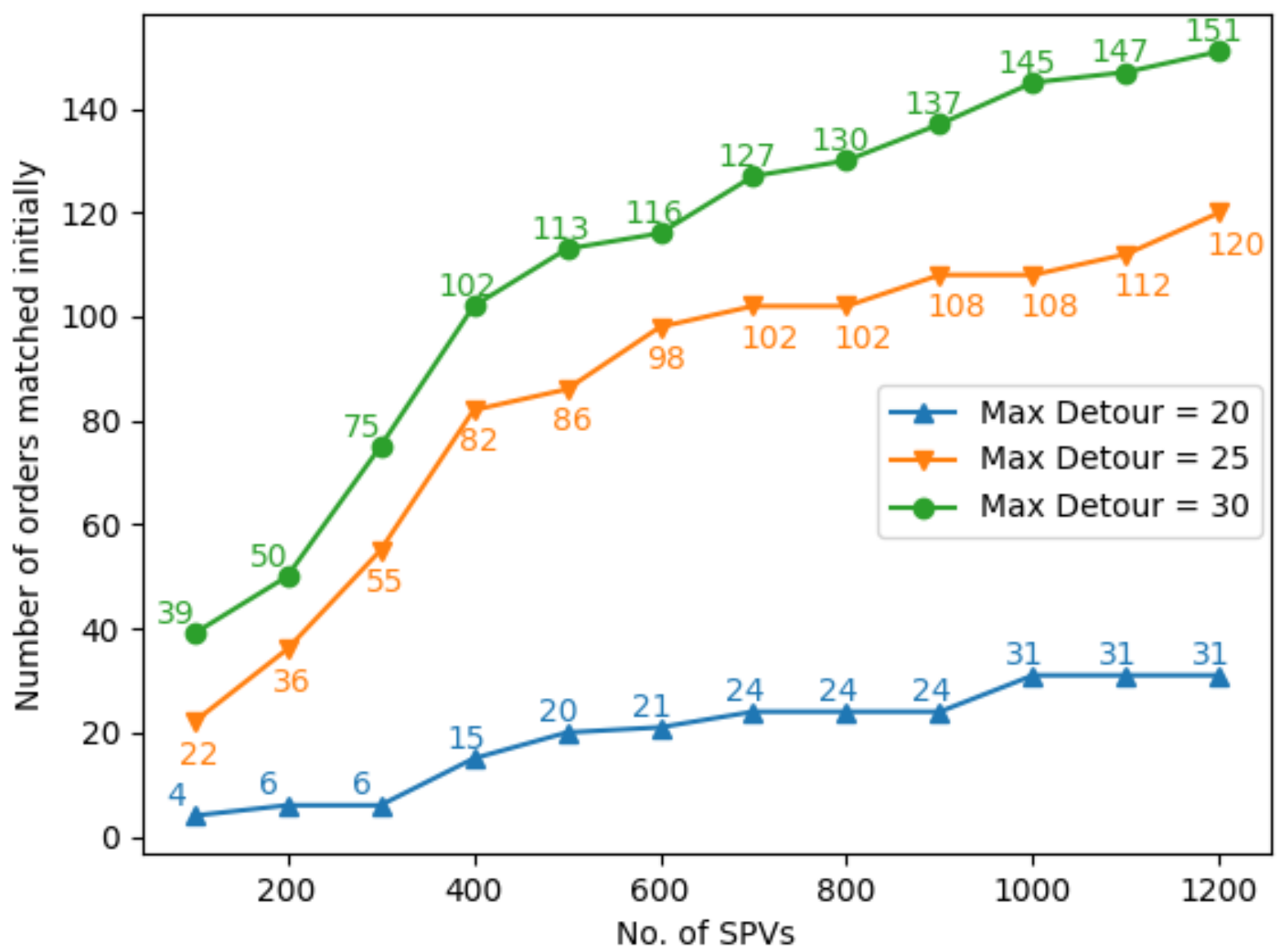

**Figure 13: Maximum PDOs that SPVs can serve**

A related parameter of interest is the percentage of SPVs that can deliver at least one PDO without violating any time-window constraints, i.e., $(no. feasible\ SPVs/total\ SPV\ number)\%$, as a function of the maximum willingness to detour. According to computational results for the Irvine case study with a 20-minute willingness to detour, only 2 ~3% of SPVs can feasibly serve a PDO. While with a 25-minute detour, the percentage increases to 18%. For the 30-min, the feasible SPV percentage is 40%. This result substantiates the finding that willingness to detour significantly impacts the potential of the CSD service. Serving a substantial number of PDOs becomes difficult without at least a 25-minute willingness to detour for SPVs.

The results in Section 6.5 show that when SPVs handle more PDOs, total VMT increases. The reason is that DV routes with over a dozen PDOs are quite efficient, whereas many SPVs serving only a couple of PDOs each must all detour to the depot for PDO pickup. Our finding is consistent with the results of Qi et al. (2018) and Rai et al. (2017). In this subsection, we extend our analysis from VMT to environmental impact, and we focus on the role of SPV willingness-to-detour.

Vehicle type substantially impacts CSD emissions (Mohri et al., 2023). This section assumes that all vehicles are gasoline-powered, all SPVs are family-size sedans or wagons, and all DVs are light-duty trucks. We analyze $CO_2$ emissions from vehicle exhaust. We obtain $CO_2$ emissions rates for different vehicle types from the Bureau of Transportation Statistics (U.S. Department of Transportation, 2023). Figure 14 to Figure 16 illustrate the total exhaust $CO_2$ (in kilograms) under different willingness-to-detour scenarios.

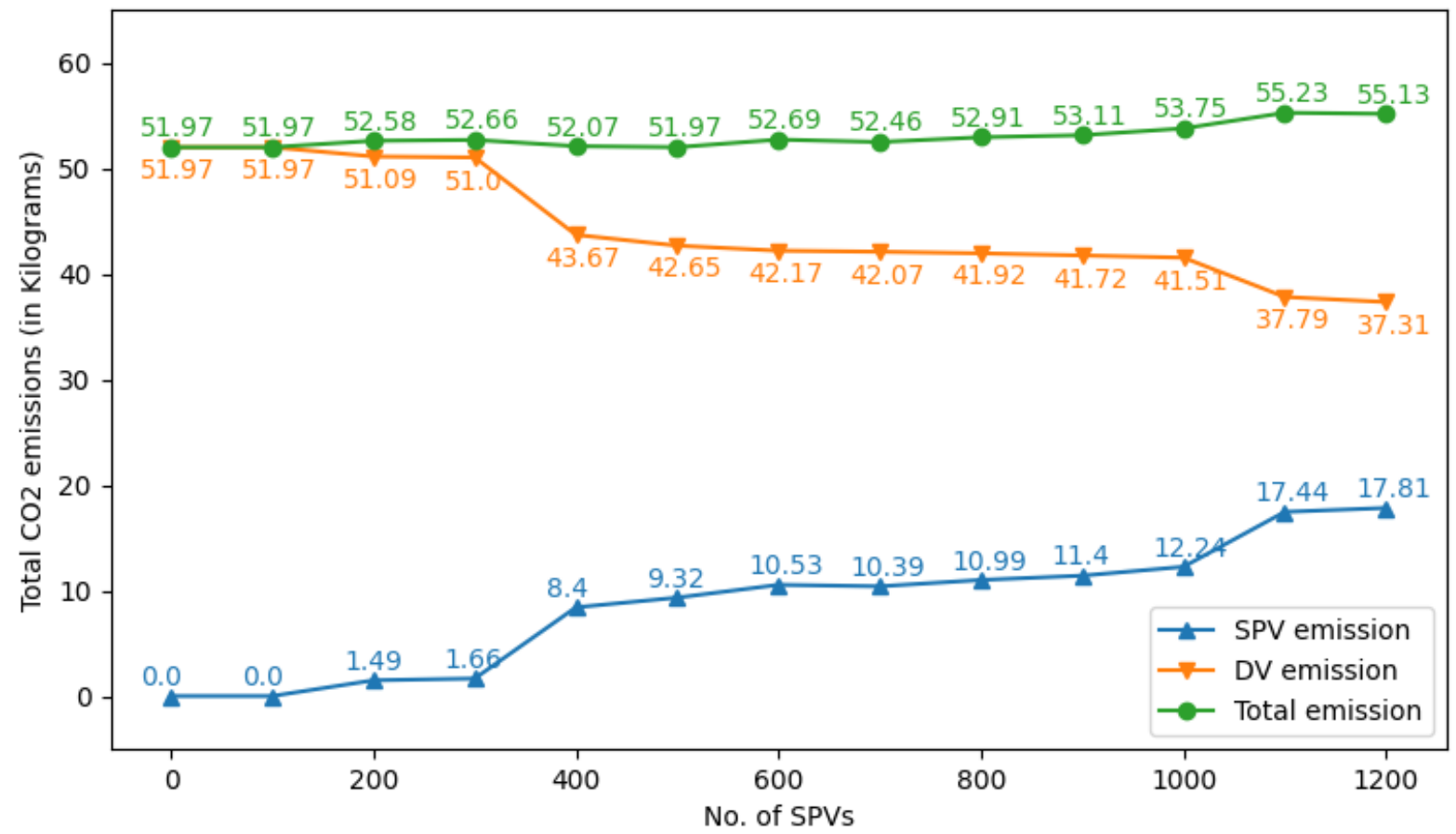

**Figure 14: $CO_2$ emissions under a 20-min maximum detour scenario**

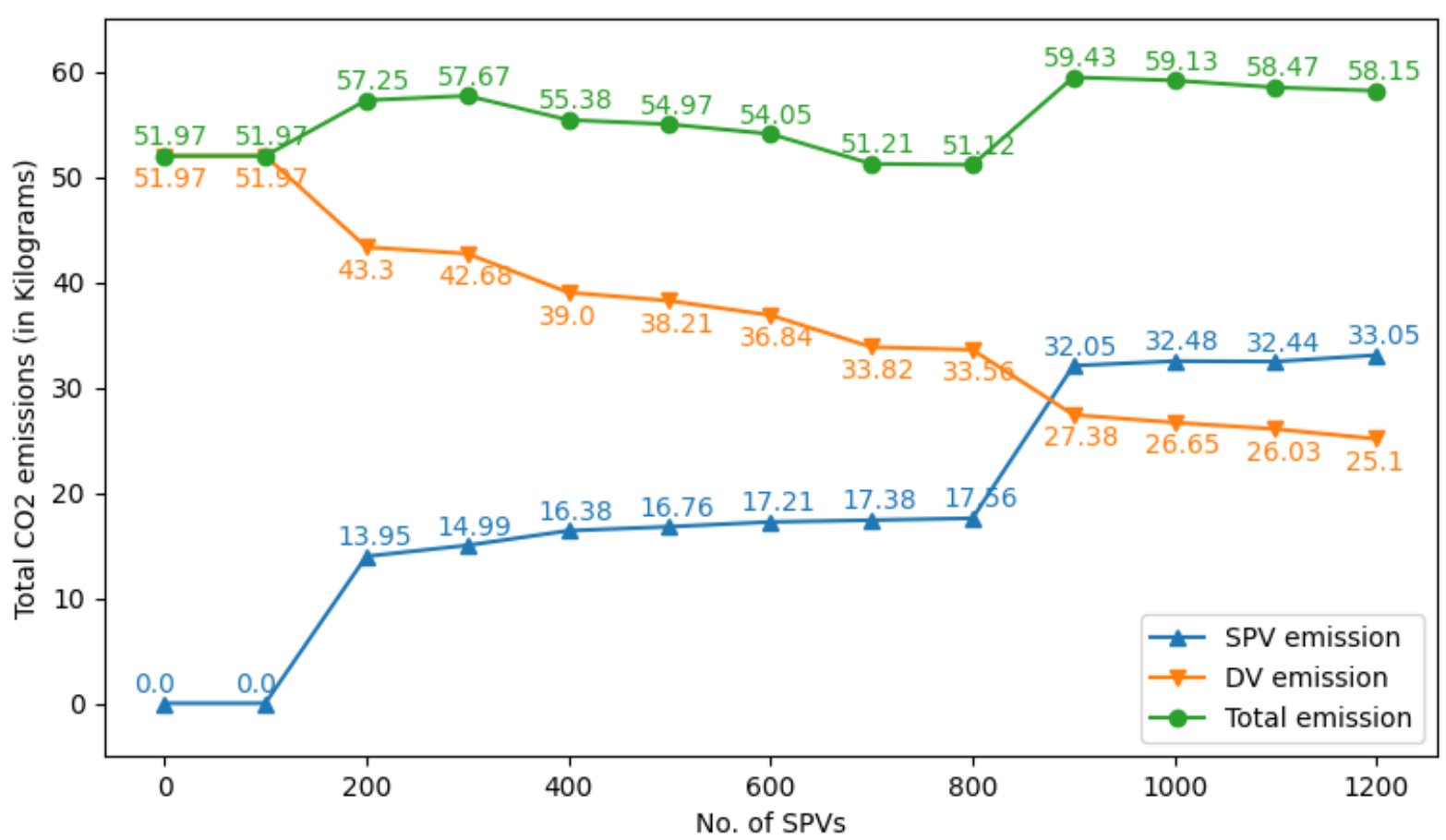

**Figure 15: $CO_2$ emissions under a 25-min maximum detour scenario**

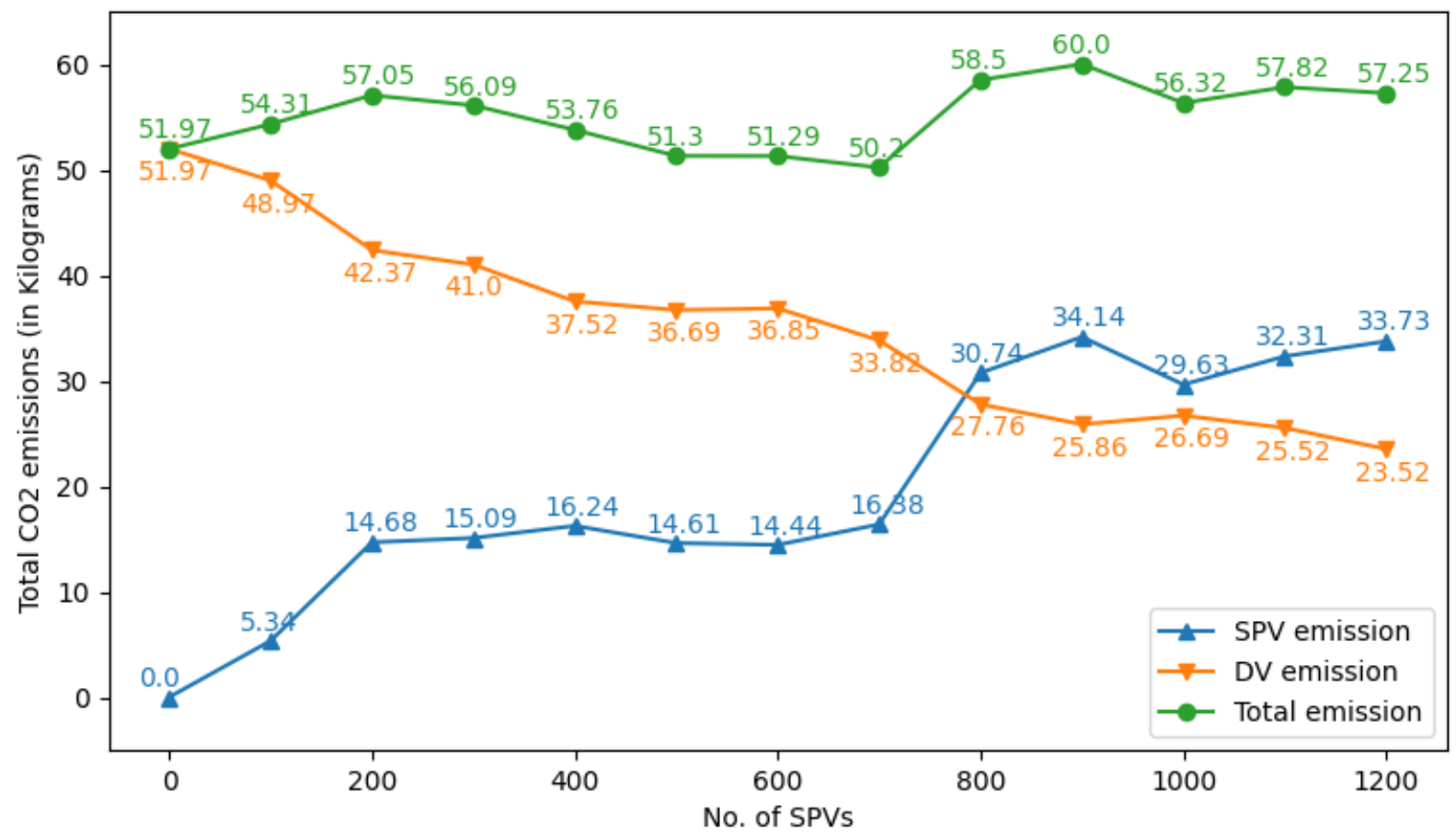

**Figure 16: $CO_2$ Emissions under a 30-min maximum detour scenario**

In all three detour scenarios, the total $CO_2$ emissions increase with the number of SPVs available and, more importantly, the number of PDOs served by SPVs. Across the three scenarios, the likelihood of matching PDOs with SPVs increases as SPV willingness-to-detour increases. Therefore, longer SPV willingness-to-detour values result in higher VMT and emissions. As such, any per-mile emissions benefit from using family-sized sedans or wagons instead of a light-duty truck cannot compensate for the increase in total VMT from SPVs.

Interestingly, however, the increase in emissions is not monotone. Using the 30-minute willingness to detour scenario as an example (Figure 16) and cross-referencing Table 6, we see that the big jumps in emissions occur when there is a significant positive jump in PDOs assigned to SPVs (and a significant negative jump in PDOs assigned to DVs).

To mitigate emissions from SPVs, CSD companies can prioritize electric SPVs when acquiring SPV drivers and assigning PDOs to SPVs.

## 6.7 Impact of Driver Rejection

In the base scenario, we assume that drivers always accept the assignment of PDOs. However, drivers may reject the assignment or not show up. Rejection may stem from dissatisfaction with the compensation or unwillingness to detour as much as is required. This section compares two scenarios with a random driver rejection rate of 10% (Scenario 1) and 20% (Scenario 2).

Algorithm 3 below describes our method to analyze the impact of SPV drivers rejecting PDOs, where $\alpha$ is the rejection probability for a PDO. After an SPV driver initially rejects a PDO, we assume the logistics company has one chance to reassign the rejected PDOs to previously unassigned SPVs. If this "second chance SPV match" is unsuccessful or the second SPV rejects a PDO, the unassigned PDOs must be delivered by DVs.

| | **Assignment with driver rejection** |
|---|---|
| *1* | **Step 1:** |
| *2* | Obtain an initial solution with the D-H algorithm |
| *3* | **Step 2:** |
| *4* | **For** $s$ in $SPV$ set: |
| *5* | **if** $random(0,1) \geq 1-\alpha$ |
| *6* | PDO $p$ rejected. Append $p$ to $PDO_{rej}$ |
| *7* | **Step 3:** |
| *8* | Rematch $PDO_{rej}$ with unassigned $SPVs$, with constraints that rule out previous rejections. |
| *9* | Repeat Step 2 for the new matchings and $PDO_{rej}$ |
| *10* | **Step 4:** |
| *11* | Run insertion algorithm to assign $PDO_{rej}$ to $DVs$. |

Figure 17 displays the total cost of each scenario as a bar plot. As expected, as the rejection percentage increases, the total cost increases. Interestingly, total cost appears to increase by 0.5-2.0% when going from 0% rejection to 20% rejection. Hence, the CSD system is naturally robust to drivers rejecting PDOs.

The line graphs in Figure 17 display the percentage differences for Scenario 1 and 2 vs. the base scenario. As the number of available SPVs increases, both "relative cost difference" lines have a sharp jump and then slowly decrease. In the 100 SPV case, the total cost differences are quite small because, with a low level of SPV participants, the number of PDOs assigned to SPVs is low. Therefore, the impact of SPV driver rejection is limited. In the 200-to-400-SPV case, the impact of rejection is the largest because the system assigns a lot of PDOs to SPVs, but when these PDOs are rejected, there are very few unassigned SPVs that can serve the rejected PDOs. In the case of 500-1200 SPVs, both lines show a decreasing trend because as the number of SPVs increases, as SPV drivers reject PDOs, other SPVs can relatively easily handle the PDOs.

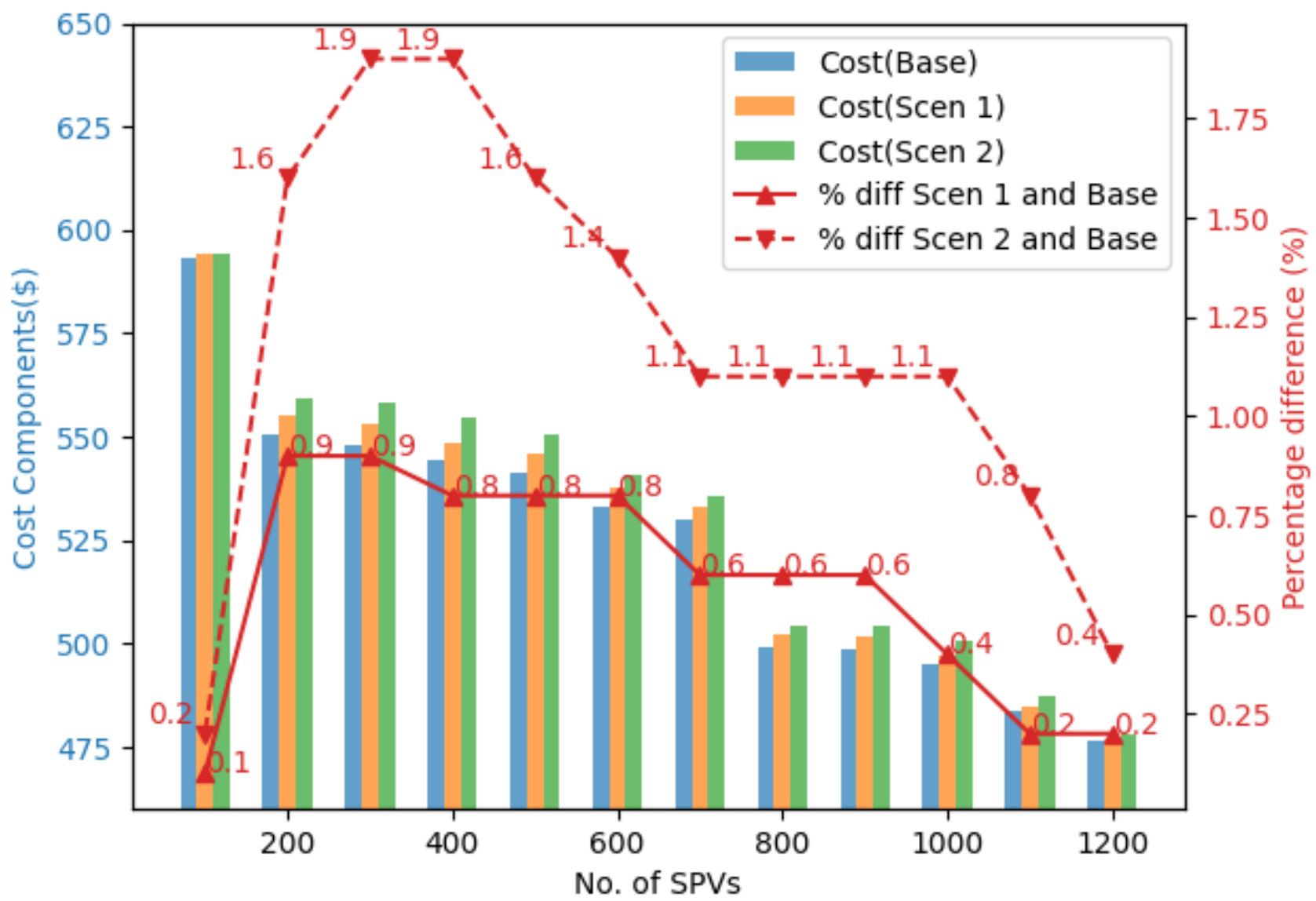

**Figure 17: Cost comparison between different driver rejection rates**

## 6.8 Managerial Insights

Section 6 includes a realistic city-scale numerical case study of a CSD system. Although the results are generated based on the City of Irvine's geographical, network, and spatial demand patterns, we believe several important findings provide transferable managerial insights.

First, the results indicate that logistics companies that employ CSD service can reduce the number of DVs they purchase and operate on a given day. By reducing the number of operating DVs, logistics companies can reduce labor costs and the fixed per-vehicle costs of storage, maintenance, and insurance. However, our results indicate that serving all PDOs using only SPVs is difficult; hence, it is hard to completely eliminate DV service.

Second, regarding externalities, our results indicate that the CSD system is likely to increase VMT and emissions relative to a DV-only system. However, when calculating emissions, we assume SPVs are sedans and DVs are light-duty tracks; we also assume both vehicle types are gas-powered. Our result contrasts with prior research on CSD systems where CD drivers start their trips from the store/depot. To mitigate emissions from SPVs, CSD companies can prioritize electric SPVs when acquiring SPV drivers and assigning PDOs to SPVs.

Third, the results also indicate that the maximum willingness-to-detour of SPV drivers has a large and highly nonlinear relationship with the cost-effectiveness of the CSD system. In our case study, there is a significant gap in performance between the 20- and 25-minute parameter values and a small gap between the 25- and 30-minute parameter values. Hence, logistics providers should thoroughly analyze the relationship between maximum willingness-to-detour and CSD costs for their service region.

# 7 Conclusion

## 7.1 Summary

This study presents new mathematical models and algorithms to address large-scale CSD problem instances. Following the literature on CD, the paper first models the CSD problem as an *MFOCVRPTW*. This formulation captures the general features of CSD. However, finding an optimal solution based on this formulation is computationally expensive. To improve efficiency, we reformulate the problem as a set partitioning problem. The alternative set partitioning formulation inspires a new solution approach called the D-H algorithm.

The novel D-H algorithm decomposes the problem by vehicle type: SPV and DV. The D-H algorithm matches PDOs to each vehicle type separately and routes each vehicle type separately in the initial stage to obtain an initial solution. After that, the algorithm switches PDOs between SPVs and DVs to improve solution quality. The switching process is an SA-inspired procedure. We compare the D-H with a commercial solver, Gurobi, for solving the CSD problem. The D-H can obtain solutions with a 1.2 % optimality gap for small-scale cases, and the D-H is much faster and more scalable than the commercial solver.

We apply the models and algorithms to the City of Irvine. The paper analyzes major factors that impact the efficiency of a CSD. The analysis includes metrics such as total cost, total VMT, and the number of PDOs delivered by SPVs. Moreover, the study presents a sensitivity analysis of various performance metrics with respect to changes in the maximum willingness to detour.

Overall, the paper significantly advances the state-of-the-art through novel, computationally efficient, and operationally effective mathematical models and associated heuristic solution algorithms. Moreover, the paper contributes to logistics and transportation practice by addressing real-world problem instances (through numerical case studies) and providing valuable managerial insights.

## 7.2 Limitation and Future Research

There are several limitations associated with our research methodology. First, we assume information about SPV drivers and PDOs is deterministic. In contrast, the real-world problem is inherently stochastic and needs to be solved online/dynamically. Future research should explicitly model uncertainty about PDOs and SPV drivers. Moreover, travel times between PDO locations are inherently uncertain. Future research can also consider stochastic travel times and account for them when generating SPV routes.

Second, while we analyze the impact of SPV driver rejection on system performance, we use a simple stochastic model with a universal rejection probability for all drivers. Future research should collect data and then estimate and calibrate a driver rejection model, wherein drivers reject assigned PDOs based on the attributes of the offer (e.g., compensation, detour distance, detour time, time-of-day). Moreover, since SPV driver rejection is likely to occur dynamically during the day, a stochastic-dynamic model is preferable for handling SPV driver rejections.

The research presented in this study also opens several additional lines of research. First, while this study assumes the central logistics operator has complete information about the origin and destination of each SPV (information that significantly impacts the fee paid to SPVs), obtaining this information in the real world may not be straightforward. Hence, developing truth-telling mechanisms/incentives to obtain this information from SPV drivers may be necessary.

To further improve the computational speed and quality of the algorithm, researchers can design additional algorithms to generate “potentially good” SPV routes or to eliminate unpromising routes. Similarly, researchers might improve the matching procedure through a learning process that can identify potentially good matches. Learning algorithms might also generate promising candidate routes for SPVs. Via learning promising candidate routes, it may be possible to reduce computational time.

Moreover, a future CD system may include connected automated vehicles owned by individuals but rented out to logistics providers during certain portions of the day. Therefore, further research could combine the usage of shared-use connected automated vehicles for people trips and PDO delivery to improve outcomes, such as reducing carbon emissions and improving vehicle utilization. Fehn et al. (2023) present a promising approach for integrating packages and people in a logistics service, where they use heuristic insertion algorithms in an agent-based simulation.

## Acknowledgments

The authors declare that we have no conflicts of interest.

We want to thank our colleague Dr. Navjyoth Sarma Jayashankar Shobha for discussing the problem and providing suggestions. We also would like to thank the editors and four anonymous reviewers for dedicating time and effort to reviewing our paper and providing insightful comments and valuable suggestions.

# Appendix A

## Pseudocode for the Budgeted Yen's Algorithm

| | **Yen's Algorithm with Budget Constraints** |
|---|---|
| *1* | $G = (V, E), source = s, target = t, budget = B$ |
| *2* | Function $\boldsymbol{BgtYenKsp}(G, s, t, B)$: |
| *3* | **Initialization**: |
| *4* | Find the shortest path between $s\ and\ t$, store it in list $A$ |
| *5* | $A = [Dijkstra(G, s, t)], A_{cost} = [SHP\ cost]$ |
| *6* | $Initialize\ a\ heap/priority\ queue\ for\ cost\ comparison$ |
| *7* | $C = heap.queue()$ |
| *8* | $BudgetFlag =$ **TRUE** |
| *9* | **While** $BudgetFlag\ is$ **TRUE**: |
| *10* | **For** $path\ in\ A$: |
| *11* | **For** $node\ in\ path, find\ spur\ nodes$: |
| *12* | $spurNode = node$ |
| *13* | $rootPath = path[:node]$ |
| *14* | $edgeRemove = [\ ]$ |
| *15* | **For** $path\ in\ A$: |
| *16* | **If** $rootPath = path[:spurNode]$ |
| *17* | $Remove\ links\ shared\ by\ rootPath$ |
| *18* | $Find\ SHP\ between\ spurNode\ to\ t, store\ as\ spurPath$ |
| *19* | **If** $no\ loop\ in\ (spurPath, rootPath)$: |
| *20* | $Combine\ rootPath\ and\ spurPath\ for\ the\ totalPath$ |
| *21* | $Calculate\ totalPathCost = rootPathCost + spurPathCost$ |
| *22* | $Store\ (total\ path, total\ path\ cost)\ in\ the\ heap\ C$ |
| *23* | **For** $edge\ in\ edgeRemoved$: |
| *24* | $Add\ back\ to\ removed\ edges\ to\ G$ |
| *25* | **If** $the\ heap\ C\ is\ not\ empty$: |
| *26* | $Get\ the\ first\ path\ in\ C\ and\ its\ cost, represented\ as\ pathAdd\ and\ pathAddCost$ |
| *27* | **If** $pathAdd\ no\ in\ A, and\ pathAddCost \leq Budget$: |
| *29* | $Append\ (pathAdd)\ to\ A\ list,$ append$(pathAddCost) to\ A_{cost}$ |
| *29* | **Elif** $pathAddCost > B$: |
| *30* | $Flip\ the\ budget\ flag, BudgetFlag =$ **FALSE** |
| *31* | **Return** $A, A_{cost}$ |

# Appendix B

## Pseudocode for the m-VRP Insertion Algorithm

| | **Insertion Algorithm for m-VRP** |
|---|---|
| *1* | **Initialization**: |
| *2* | $P_{dv} = the\ set\ of\ unrouted\ package\ locations$ |
| *3* | $R = the\ set\ of\ routes$; Including an empty route |
| *4* | **While** $P_{dv} \neq \emptyset$: |
| *5* | Minimum insertion cost $c^* = +\infty$ |
| *6* | Node to insert $u^* = None$ |
| *7* | Link to insert $(i,j)^* = None$ |
| *8* | Route to insert $r^* = None$ |
| *9* | **For** all unserved nodes $u \in P_{dv}$: |
| *10* | **For** all vehicle routes $r \in R$: |
| *11* | **If** $quantity[p] + load[r] \leq Capacity\ q_k$: |
| *12* | **For** $all\ link\ (i,j) \in r, i,j\ are\ adjacent\ nodes$: |
| *13* | **Do** $c_{ins} = c_{i,u} + c_{u,j} - c_{i,j}$ |
| *14* | **If** $c_{ins} < c^*$, time window fits, and Route duration ≤ Max duration: |
| *15* | $u^* = u;\ c^* = c;$ |
| *16* | $(i,j)^* = (i,j); r^* = r;$ |
| *17* | **End If** |
| *18* | **End For** |
| *19* | **End If** |
| *20* | **End For** |
| *21* | **End For** |
| *22* | Insert $u^*\ to\ (i,j)^*\ in\ r^*$, update $r^*$. |
| *23* | $P_{dv} \setminus u^*$ |
| *24* | **End While** |
| *25* | **Return** r$oute\ set\ R$ |

# Appendix C

## Pseudocode for a Large Neighborhood Search (LNS) Algorithm

| | **Large Neighborhood Search (LNS)** |
|---|---|
| *1* | **Initialization**: |
| *2* | Input feasible solution $\boldsymbol{x}^{\boldsymbol{init}}, \boldsymbol{\Theta}^{\boldsymbol{init}}$ |
| *3* | $\boldsymbol{x}^{\boldsymbol{best}} = \boldsymbol{x}^{\boldsymbol{init}}, \boldsymbol{\Theta}^{\boldsymbol{best}} = \boldsymbol{\Theta}^{\boldsymbol{init}}$ |
| *4* | **While** iteration$(t) <$ Taget $(220)$ |
| *5* | $t += 1$ |
| *6* | **Destruction**: |
| *7* | Randomly choose 30% of the matching for rematching |
| *8* | **Reconstruction:** |
| *9* | Follow Appendix A for the SPV and PDO assignment |
| *10* | Follow Appendix B Insertion Algorithm for DV and PDO Assignment |
| *11* | Obtain new solution $\boldsymbol{x}^{\boldsymbol{t}}, \boldsymbol{\Theta}^{\boldsymbol{t}}$ |
| *12* | **If** $\boldsymbol{\Theta}^{t} < \boldsymbol{\Theta}^{\mathbf{best}}$: |
| *13* | $\boldsymbol{x}^{\boldsymbol{best}} = \boldsymbol{x}^{\boldsymbol{t}}, \boldsymbol{\Theta}^{\boldsymbol{best}} = \boldsymbol{\Theta}^{\boldsymbol{t}}$ |
| *14* | **Elif** $\boldsymbol{\Theta}^{t} \geq \boldsymbol{\Theta}^{\mathbf{best}}$: |
| *15* | $Test = \exp\left(-\frac{\Theta^{t} - \Theta^{best}}{Target}\right)$ |
| *16* | **If** $Test > Rand(0, 1)$: |
| *17* | $\boldsymbol{x}^{\boldsymbol{best}} = \boldsymbol{x}^{\boldsymbol{t}}, \boldsymbol{\Theta}^{\boldsymbol{best}} = \boldsymbol{\Theta}^{\boldsymbol{t}}$ |
| *18* | **Else**: |
| *19* | *Reject* |
| *20* | **End If** |
| *21* | **End If** |
| *22* | **End While** |
| *23* | **Return** $\boldsymbol{x}^{\boldsymbol{best}}, \boldsymbol{\Theta}^{\boldsymbol{best}}$ |